\documentclass[12pt]{article}
\pdfoutput=1
\usepackage{amsmath,amssymb,amsthm,bm,graphicx,float,array,multirow,multicol,rotfloat,caption,subcaption,hyperref,cleveref,enumerate,geometry,mathdots,adjustbox,booktabs,parskip,mathtools,tikz,tikz-cd,pdflscape,csquotes,lscape,rotating,empheq,mathrsfs}
\usepackage[para]{threeparttable}
\usepackage[all]{xy}
\usepackage[normalem]{ulem}
\usepackage[numbers,sort&compress]{natbib}

\usetikzlibrary{positioning}
\numberwithin{equation}{section}
\numberwithin{figure}{section}
\allowdisplaybreaks
\makeatletter

\usetikzlibrary{arrows,shapes.misc}
\tikzset{gauge/.style={rounded rectangle, draw=black!100,dashed, thick, minimum size=5mm},d2/.style={rounded rectangle, draw=white!100, thick, minimum size=5mm},flavor/.style={rectangle, draw=black!100, thick, minimum size=5mm},gaugeN1/.style={rounded rectangle, draw=black!100, thick, minimum size=5mm}}

\tikzset{
node/.style={rounded rectangle, thick, draw=black!100,fill=white!100,  minimum size=6mm, inner sep=3pt},
sonode/.style={circle, thick, draw=black!100,fill=red!100,  minimum size=3mm, inner sep=0pt},
spnode/.style={circle, thick, draw=black!100,fill=blue!100,  minimum size=3mm, inner sep=0pt},
fnode/.style={rectangle, thick, draw=black!100,fill=white!100,  minimum size=3mm, inner sep=0pt},
tnode/.style={rounded rectangle, outer sep=0pt, thick, minimum size=5mm}
}

\theoremstyle{plain}
\newtheorem*{thm*}{Theorem}

\theoremstyle{definition}

\newtheorem*{defn*}{Definition}

\makeatother

\newcommand{\trace}{\operatorname{Tr}}
\newcommand{\CN}{\mathcal{N}}

\begin{document}

\begin{titlepage}
\vspace*{-3cm} 
\begin{flushright}
{\tt CALT-TH-2022-035}\\
{\tt DESY-22-152}\\
\end{flushright}
\begin{center}
\vspace{2cm}
{\LARGE\bfseries Operator spectroscopy for 4d SCFTs\\ with $a = c$\\}
\vspace{1.2cm}

{\large
Monica Jinwoo Kang$^1$, Craig Lawrie$^2$, Ki-Hong Lee$^3$, and Jaewon Song$^3$\\}
\vspace{.7cm}
{ $^1$ Walter Burke Institute for Theoretical Physics, California Institute of Technology}\par
{Pasadena, CA 91125, U.S.A.}\par
\vspace{.2cm}
{ $^2$ Deutsches Elektronen-Synchrotron DESY,}\par
{Notkestr.~85, 22607 Hamburg, Germany}\par
\vspace{.2cm}
{ $^3$ Department of Physics, Korea Advanced Institute of Science and Technology}\par
{Daejeon 34141, Republic of Korea}\par
\vspace{.2cm}
\vspace{.3cm}

\scalebox{.8}{\tt monica@caltech.edu, craig.lawrie1729@gmail.com, khlee11812@gmail.com, jaewon.song@kaist.ac.kr}\par
\vspace{1.2cm}
\textbf{Abstract}
\end{center}

We study a rich set of four-dimensional superconformal field theories (SCFTs) with both central charges identical: $a = c$. These are constructed via the diagonal $\mathcal{N}=2$ or $\mathcal{N}=1$ gauging of the flavor symmetry $G$ of a collection of $\mathcal{N}=2$ Argyres--Douglas theories of type $\mathcal{D}_p(G)$, with or without adjoint chiral multiplets, in \href{https://arxiv.org/abs/2106.12579}{\tt 2106.12579} and \href{https://arxiv.org/abs/2111.12092}{\tt 2111.12092}. We compute superconformal indices of some theories where the rank of $G$ is low, performing a refined test for unitarity, and further determine the relevant and marginal operator content in detail. We find that most of these theories flow to interacting SCFTs with $a=c$ in the infrared. 

\vfill 
\end{titlepage}

\tableofcontents

\newpage

\section{Introduction}

It is always a fruitful endeavor to study non-perturbative aspects of four-dimensional superconformal field theories. From the non-perturbative perspective, it can be challenging to determine the local operator spectrum, including subtle relations between operators and the presence of renormalization group flows between different SCFTs. 
A systematic way to tackle this problem is via utilizing the superconformal index \cite{Kinney:2005ej, Romelsberger:2005eg}. With this refined tool in hand, we study the operator contents of a variety of 4d $\mathcal{N}=1$ and $\mathcal{N}=2$ superconformal field theories with $a=c$.

In fact, a large class of 4d $\mathcal{N}=2$ SCFTs with identical central charges, $a=c$, are studied in \cite{Kang:2021lic} via gauging the common flavor symmetry $G$ of a collection of $\mathcal{D}_{p_i}(G)$ Argyres--Douglas theories. In a similar fashion, via gauging the common flavor symmetry in the $\mathcal{N}=1$ sense, this construction has been further expanded to construct 4d $\mathcal{N}=1$ SCFTs with $a=c$ in \cite{Kang:2021ccs}.\footnote{Further 4d $\mathcal{N}=2$ SCFTs with $a=c$, arising from the class $\mathcal{S}$ perspective \cite{Gaiotto:2009we,Gaiotto:2009hg}, have recently appeared in \cite{Kang:2022zsl}.} We find that almost all asymptotically-free or conformal gaugings, potentially with the inclusion of adjoint-valued chiral matter, lead to 4d $\mathcal{N}=1$ SCFTs with $a = c$, if all of the $p_i$ are coprime with the dual Coxeter number of $G$ ($h_G^\vee$). To verify that these infrared SCFTs are indeed unitary interacting SCFTs, we must determine that there exists a non-anomalous superconformal R-symmetry via the principle of $a$-maximization \cite{Intriligator:2003jj}, and further that along the flow into the infrared there are no operators that become free and give rise to such a decoupled sector. Since free theories do not have $a = c$, their presence indicates that nor would the interacting sector. In \cite{Kang:2021ccs}, it is confirmed that the Coulomb branch operators of each $\mathcal{D}_{p_i}(G)$ theory and the moment map operators do not cross the unitarity bound during the flow. In this paper, we do a more refined check of unitarity of the 4d $\mathcal{N}=1$ theories by determining their full superconformal indices, for the cases where the gauge group is of sufficiently low rank.

In order to show that the 4d $\mathcal{N}=1$ SCFTs that we obtain in this manner truly are unitary interacting SCFTs with $a = c$, we need a further check that goes beyond the operators in the chiral ring. Certain unitarity constraints are not directly visible from the operators in the chiral ring, but are reflected in the superconformal index \cite{Beem:2012yn, Evtikhiev:2017heo, Maruyoshi:2018nod}. We perform this vital check in cases where the gauge group is low rank by computing the full $\mathcal{N}=1$ superconformal index. This relies on the known expressions for the superconformal indices of the $\mathcal{D}_p(G)$ theories, for certain specific choices of $p$ and $G$. In addition to determining that the theory is interacting, the index also provides a wealth of information about the operator content of the theory. When all fugacities are turned on, we can read off the precise details of the relevant and marginal operators of the $a = c$ theories. We refer to this process of determining the operator content from the index as ``operator spectroscopy.'' We compute the indices for gaugings with both rational and irrational $R$-charges, including some of the $\mathcal{N}=2$ $\widehat{\Gamma}(G)$ theories, and also for theories with additional adjoint chiral multiplets. In all cases, we find that there does not exist any unitarity-violating term in the superconformal indices. This procedure of operator spectroscopy enables us to determine interesting properties of the theories, such as the structure of their conformal manifolds, and the superpotential deformations that may trigger a flow to a new infrared SCFT. Intriguingly, we find that many of these deformations preserve the $a = c$ property, and we will explore the landscape of such deformations in the upcoming paper \cite{LANDSCAPE}.

In order to compute the superconformal indices for the $\CN=1$ SCFTs that we consider, we need to know the superconformal indices of the individual $\mathcal{D}_p(G)$ theories. Unfortunately, the full superconformal indices for $\mathcal{D}_p(G)$ theories are unavailable in general. However, for the $\mathcal{D}_2(SU(3))$ theories and the $\mathcal{D}_p(SU(2))=(A_1, D_p)$ theories, there are known $\mathcal{N}=1$ UV Lagrangian theories that have supersymmetry-enhancing flows to those $\mathcal{N}=2$ SCFTs \cite{Maruyoshi:2016aim, Maruyoshi:2016tqk, Agarwal:2016pjo, Agarwal:2017roi, Benvenuti:2017bpg}. From this ``Lagrangian description,'' the superconformal indices can be computed.\footnote{The Schur or Macdonald limit of the indices for a larger subset of $\mathcal{D}_p(G)$ theories is available in \cite{Buican:2015ina, Buican:2015tda, Buican:2017uka,Cordova:2015nma, Beem:2020pry, Foda:2019guo, Watanabe:2019ssf, Song:2015wta, Song:2016yfd, Agarwal:2019crm, Song:2017oew, Xie:2019zlb}.} As such, in this paper we focus on SCFTs that are constructed via a diagonal $\mathcal{N}=1$ or $\mathcal{N}=2$ gauging of the common flavor symmetry of a collection of $\mathcal{D}_2(SU(3))$ and $\mathcal{D}_p(SU(2))$ theories.

The structure of the rest of the paper is as follows. 
In Section \ref{sec:aeqc}, we start by reviewing the construction of the $\widehat{\Gamma}(G)$ theories from \cite{Kang:2021lic} and the extension to the $\mathcal{N}=1$ SCFTs with $a=c$ that were discussed in \cite{Kang:2021ccs}; these are the theories we explore throughout this paper. In Section \ref{sec:unitarity}, we introduce the superconformal index and explain how the superconformal index can detect the existence of non-unitary operators in the spectrum of the putative infrared theory. We state in Section \ref{sec:indexDpG} the superconformal indices of the building block $\mathcal{D}_p(G)$ theories that have been computed in the literature. We combine the $\mathcal{D}_p(G)$ indices to study the $\widehat{D}_4(SU(3))$ and $\widehat{E}_6(SU(2))$ SCFTs in Section \ref{sec:indexD4E6}; this allows us to determine that there are no non-unitary operators and to read off the exact operator content for low values of the scaling dimension.
In Section \ref{sec:conformal}, we study the conformal manifolds for the $\mathcal{N}=1$ SCFTs with $a=c$ obtained by conformal gauging of a collection of $\mathcal{D}_p(G)$ theories.
In the subsequent four sections, we apply the technique of operator spectroscopy to determine the operator content of a variety of $\mathcal{N}=1$ SCFTs studied in \cite{Kang:2021ccs}. We study theories built out of $\mathcal{D}_2(SU(3))$ Argyres--Douglas SCFTs in Section \ref{sec:multipleD2SU3}; we explore theories built out of the $\mathcal{D}_3(SU(2))$ theories in Section \ref{sec:multipleD3SU2}; in Section \ref{sec:indexirrational}, we study gaugings involving $\mathcal{D}_5(SU(2))$, together with the previously considered building blocks; in Section \ref{sec:indexadjoints}, we study theories where the gauging also involves additional adjoint-valued chiral multiplets. To conclude, in Section \ref{sec:summary}, we provide tables of the relevant and marginal operator content that we determine from the technique of operator spectroscopy throughout this paper; we further provide some future applications of this knowledge of the operator content. We list in Appendix \ref{app:indicesplusplus} the fully flavor-refined indices for the theories we study in this paper.

\section{SCFTs with \texorpdfstring{\boldmath{$a=c$}}{a=c} from \texorpdfstring{\boldmath{$\mathcal{N}=2$}}{N=2} and \texorpdfstring{\boldmath{$\mathcal{N}=1$}}{N=1} gaugings}
\label{sec:aeqc}

In this paper, we exemplify the technique of operator spectroscopy in the context of the $\mathcal{N}=2$ and $\mathcal{N}=1$ SCFTs with $a = c$ that were discussed in \cite{Kang:2021ccs,Kang:2021lic}.\footnote{4d SCFTs with $a=c$ have also recently been discussed in \cite{Carta:2021whq,Closset:2021lwy,Kang:2022zsl}.} The 4d $\mathcal{N}=2$ SCFTs of interest are constructed out of the following building blocks: the Argyres--Douglas $\mathcal{D}_p(G)$ theories \cite{Cecotti:2012jx,Xie:2012hs,Cecotti:2013lda,Wang:2015mra}, the minimal $(G, G)$ conformal matter theories \cite{DelZotto:2014hpa,Ohmori:2015pua}, and $\mathcal{N}=2$ vector multiplets. It was studied in \cite{Kang:2021ccs} under what circumstances can one take a set of such building blocks and gauge together all of the $G$ flavor symmetries with $\mathcal{N}=2$ vector multiplets such that one obtains a superconformal field theory. 

To analyze this question, it is important to understand the physical properties of these building blocks. The $\mathcal{D}_p(G)$ theories can be obtained from the class $\mathcal{S}$ perspective as compactifications of the 6d $(2,0)$ SCFT of type $G$ on a sphere with a regular maximal puncture and an irregular puncture. The regular puncture provides the theory with a flavor symmetry $G$ with flavor central charge
\begin{equation}\label{eqn:ADlevel}
    k_G^\text{AD} = \frac{2(p-1)}{p}h_G^\vee \,,
\end{equation}
where $h_G^\vee$ is the dual Coxeter number of $G$. The irregular puncture may also provide an additional flavor symmetry factor, depending on $p$ and $G$. We have summarized when these extra flavor symmetries occur in Table \ref{tbl:extraflv}. 

\begin{table}[H]
    \begin{threeparttable}
    \centering
    \renewcommand{\arraystretch}{1.2}
    \centering
    $\begin{array}{c|ccccc}
        \toprule
        G & SU(N) & SO(2N) & E_6 & E_7 & E_8 \\
        \midrule
        \text{No additional symmetry} & (p, N) = 1 & p \notin 2\mathbb{Z}_+ & p \notin 3\mathbb{Z}_+ & p \notin 2\mathbb{Z}_+ & p \notin 30\mathbb{Z}_+
        \\\bottomrule
    \end{array}$
    \end{threeparttable}
    \caption{The conditions required to be satisfied if the irregular puncture of the $\mathcal{D}_p(G)$ theory does not contribute any flavor symmetry.}
    \label{tbl:extraflv}
\end{table}

The minimal $(G, G)$ conformal matter theories can be obtained from the class $\mathcal{S}$ perspective by starting from the 6d $(2,0)$ SCFT of type $G$ and compactifying on a sphere with two regular maximal punctures and one simple puncture.\footnote{These theories are referred to as conformal matter as they, and their descendants via Higgs branch renormalization group flows, also arise from the 6d $(1,0)$ theories known as minimal $(G, G)$ conformal matter \cite{DelZotto:2014hpa} compactified on a torus. See \cite{DelZotto:2015rca,Ohmori:2015pia,Ohmori:2015pua,Baume:2021qho} and references therein for more details.} The regular punctures each contribute a flavor symmetry $G$; both have the same flavor central charge
\begin{equation}\label{eqn:CMlevel}
    k_G^\text{L}=k_G^\text{R}=2h_G^\vee \,.
\end{equation}
There are additional Abelian flavor symmetries when $G = SU(N)$, and an additional $SU(2)$ flavor symmetry when $G = SU(2)$, though we shall not be concerned with those in this paper.

To obtain a conformal field theory, it is necessary for the one-loop $\beta$-function of the gauge coupling for each introduced $\mathcal{N}=2$ vector multiplet to vanish: $\beta_G = 0$. We assume that an $\mathcal{N}=2$ vector multiplet is introduced that gauges the flavor symmetry of $n$ copies of $\mathcal{D}_{p_i}(G)$ and $m$ factors of the $(G, G)$ minimal conformal matters which provide links between two gauge nodes or from a gauge node to itself. Then the condition on the vanishing of the one-loop $\beta$-function is, using the expressions for the flavor central charges in equations \eqref{eqn:ADlevel} and \eqref{eqn:CMlevel}, 
\begin{equation}\label{eqn:N2beta}
    \beta_G = 0 \quad \Longleftrightarrow \quad \sum_{i=1}^n \frac{2(p_i - 1)}{p_i}h_G^\vee + 2mh_G^\vee = 4 h_G^\vee \,, 
\end{equation}
where the RHS is the contribution from the introduced vector multiplet.
It was shown in \cite{Kang:2021ccs} that there are only six solutions satisfying this equality for finite $p_i$. The first four solutions involve no copies of conformal matter, i.e., $m = 0$, and they can be written as the following quivers, respectively, corresponding to $\widehat{D}_4(G)$, $\widehat{E}_6(G)$, $\widehat{E}_7(G)$, and $\widehat{E}_8(G)$ theories:
\begin{equation}\label{eqn:GammahatG}
\begin{gathered} \widehat{D}_4(G):\ 
  \begin{gathered}
  \begin{tikzpicture}
    \node[tnode] (N1) {$\mathcal{D}_2(G)$};
    \node[node] (N2) [right=4mm of N1] {$G$};
    \node[tnode] (N3) [right=4mm of N2] {$\mathcal{D}_2(G)$};
    \node[tnode] (Nu) [above=4mm of N2] {$\mathcal{D}_2(G)$};
    \node[tnode] (Nl) [below=4mm of N2] {$\mathcal{D}_2(G)$};
    \draw (N1.east) -- (N2.west);
    \draw (N2.east) -- (N3.west);    
    \draw (N2.north) -- (Nu.south);
    \draw (N2.south) -- (Nl.north);
  \end{tikzpicture}
  \end{gathered} \,, \qquad\quad\widehat{E}_6(G):\ 
  \begin{gathered}
  \begin{tikzpicture}
    \node[tnode] (N1) {$\mathcal{D}_3(G)$};
    \node[node] (N2) [right=4mm of N1] {$G$};
    \node[tnode] (N3) [right=4mm of N2] {$\mathcal{D}_3(G)$};
    \node[tnode] (Nu) [above=4mm of N2] {$\mathcal{D}_3(G)$};
    \draw (N1.east) -- (N2.west);
    \draw (N2.east) -- (N3.west);    
    \draw (N2.north) -- (Nu.south);
  \end{tikzpicture}
  \end{gathered} \,, \\\widehat{E}_7(G):\ 
  \begin{gathered}
  \begin{tikzpicture}
    \node[tnode] (N1) {$\mathcal{D}_4(G)$};
    \node[node] (N2) [right=4mm of N1] {$G$};
    \node[tnode] (N3) [right=4mm of N2] {$\mathcal{D}_4(G)$};
    \node[tnode] (Nu) [above=4mm of N2] {$\mathcal{D}_2(G)$};
    \draw (N1.east) -- (N2.west);
    \draw (N2.east) -- (N3.west);    
    \draw (N2.north) -- (Nu.south);
  \end{tikzpicture}
  \end{gathered} \,, \qquad\quad\widehat{E}_8(G):\ 
  \begin{gathered}
  \begin{tikzpicture}
    \node[tnode] (N1) {$\mathcal{D}_3(G)$};
    \node[node] (N2) [right=4mm of N1] {$G$};
    \node[tnode] (N3) [right=4mm of N2] {$\mathcal{D}_6(G)$};
    \node[tnode] (Nu) [above=4mm of N2] {$\mathcal{D}_2(G)$};
    \draw (N1.east) -- (N2.west);
    \draw (N2.east) -- (N3.west);    
    \draw (N2.north) -- (Nu.south);
  \end{tikzpicture}
  \end{gathered} \,.
\end{gathered}
\end{equation}
When conformal matter is included we find that $m \leq 2$, and the only two options are:
\begin{equation}\label{eqn:GammahatGm}
  \begin{gathered}
  \begin{tikzpicture}
    \node[tnode] (N1) {$\mathcal{D}_2(G)$};
    \node[node] (N2) [right=4mm of N1] {$G$};
    \node[tnode] (N3) [right=4mm of N2] {};
    \node[tnode] (Nu) [above=4mm of N2] {$\mathcal{D}_2(G)$};
    \draw (N1.east) -- (N2.west);
    \draw (N2.east) -- (N3.west);    
    \draw (N2.north) -- (Nu.south);
  \end{tikzpicture}
  \end{gathered} \,, \qquad\quad
  \begin{gathered}
  \begin{tikzpicture}
    \node[tnode] (N1) {};
    \node[node] (N2) [right=4mm of N1] {$G$};
    \node[tnode] (N3) [right=4mm of N2] {};
    \draw (N1.east) -- (N2.west);
    \draw (N2.east) -- (N3.west);    
  \end{tikzpicture}
  \end{gathered} \,.
\end{equation}
Here, a solid line that is not connected to anything on one side represents a $(G, G)$ conformal matter theory where only one of the $G$ flavor symmetries has been gauged. To determine all possibilities for superconformal theories that can be obtained by gauging together all of the $G$ flavor symmetries of a collection of such building blocks, we determine how each of these gauge nodes can be adjoined. Clearly the configurations in equation \eqref{eqn:GammahatG} cannot be connected to any other gauge node, and the configurations with an open conformal matter link can only be connected together in the following two ways
\begin{equation}
    \begin{gathered}
  \begin{tikzpicture}
    \node[tnode] (N1) {$\mathcal{D}_2(G)$};
    \node[node] (N2) [right=4mm of N1] {$G$};
    \node[tnode] (N3) [right=4mm of N2] {$\cdots$};
    \node[node] (N4) [right=4mm of N3] {$G$};
    \node[tnode] (N5) [right=4mm of N4] {$\mathcal{D}_2(G)$};
    \node[tnode] (Nu) [above=4mm of N2] {$\mathcal{D}_2(G)$};
    \node[tnode] (Nu2) [above=4mm of N4] {$\mathcal{D}_2(G)$};
    \draw (N1.east) -- (N2.west);
    \draw (N2.east) -- (N3.west);   
    \draw (N3.east) -- (N4.west); 
    \draw (N4.east) -- (N5.west); 
    \draw (N2.north) -- (Nu.south);
    \draw (N4.north) -- (Nu2.south);
  \end{tikzpicture}
  \end{gathered} \,, \qquad\quad
  \begin{gathered}
  \begin{tikzpicture}
    \node[node] (N1) {$G$};
    \node[tnode] (N2) [right=4mm of N1] {$\cdots$};
    \node[node] (N3) [right=4mm of N2] {$G$};
    \node[node] (Nu) [above=4mm of N2] {$G$};
    \draw (N1.east) -- (N2.west);
    \draw (N2.east) -- (N3.west); 
    \draw (N1.45) -- (Nu.225);
    \draw (N3.135) -- (Nu.315);
  \end{tikzpicture}
  \end{gathered} \,.
\end{equation}
We refer to these theories as $\widehat{D}_{N+3}(G)$ and $\widehat{A}_{N-1}(G)$, respectively, where $N$ is the number of gauge nodes in the quiver. Thus, we can see that superconformal $\mathcal{N}=2$ quiver gauge theories formed by gauging together copies of $\mathcal{D}_p(G)$ and $(G, G)$ conformal matter have an ADE-type classification, and we label them collectively as $\widehat{\Gamma}(G)$ \cite{Kang:2021ccs}. 

It turns out that a subset of the $\widehat{\Gamma}(G)$ theories have identical central charges: $a = c$. These cases occur when there are no $(G, G)$ conformal matter theories involved in the gauging -- these are the configurations that were depicted in equation \eqref{eqn:GammahatG}; i.e., when $\Gamma = D_4$, $E_6$, $E_7$, or $E_8$.\footnote{These specific families of theories are sometimes known as the elliptic $G$-models \cite{Cecotti:2011rv}; for some of these theories, aspects have been explored in \cite{Buican:2016arp,Buican:2020moo,DelZotto:2015rca,Closset:2020afy,Cecotti:2013lda,Closset:2021lwy}.} When $\gcd(p_i, h_G^\vee) = 1$ then the central charges of the $\mathcal{D}_{p_i}(G)$ building block become
\begin{equation}
a_i =\frac{1}{48}\frac{(4p_i-1)(p_i-1)}{p_i}\operatorname{dim}(G)\,, \quad 
c_i =\frac{1}{12}(p_i-1)\operatorname{dim}(G)\,.
\end{equation}
In these circumstances, it is easy to see that the difference of the central charges of the gauged theories are
\begin{equation}
    \begin{aligned}
        48(c - a) &= -2\operatorname{dim}(G) + \sum_{i}\bigg( 4(p_i - 1) -  \frac{(4p_i-1)(p_i-1)}{p_i} \bigg)\operatorname{dim}(G) \cr 
        &= \operatorname{dim}(G)\bigg( -2 + \sum_{i} \frac{(p_i - 1)}{p_i} \bigg) = 0 \,,
    \end{aligned}
\end{equation}
where the first term comes from the vector multiplet and the last equality follows by application of equation \eqref{eqn:N2beta}. We note that the gcd-condition between the $p_i$ and $h_G^\vee$ can be written more succinctly as
\begin{equation}
    \prod_i \gcd(p_i, h_G^\vee) = 1 \quad \Longleftrightarrow \quad \gcd(\alpha_\Gamma, h_G^\vee) = 1 \,,
\end{equation}
where $\alpha_\Gamma$ is the largest comark associated to the Dynkin diagram $\Gamma$. These $a = c$ theories have an interesting connection, in the Schur sector, to $\mathcal{N}=4$ super-Yang--Mills, which was explored in \cite{Kang:2021ccs} (see also \cite{Buican:2020moo,Honda:2022hvy,Huang:2022bry,Hatsuda:2022xdv}), and to which we refer the reader for more details.

In \cite{Kang:2021lic}, the authors considered an extension of the above analysis to gauging the building blocks via an $\mathcal{N}=1$ vector multiplet instead of an $\mathcal{N}=2$ vector multiplet. In this case, we can also consider an additional building block: chiral multiplets transforming in a representation $\bm{R}$ of $G$. The condition for the coupling of the introduced gauge node to be either asymptotically-free or conformal is, schematically, 
\begin{equation}\label{eqn:betaone}
    \beta_G \leq 0 \quad \Longleftrightarrow \quad \sum_{\mathcal{D}_{p_i}(G)s} \frac{2(p_i - 1)}{p_i}h_G^\vee + \sum_{\substack{\text{conformal} \\ \text{matters}}}2h_G^\vee + \sum_{\text{chirals}} I(\bm{R}) \leq 6h_G^\vee \,,
\end{equation}
where the sums are over the different types of building blocks connected to that $\mathcal{N}=1$ gauge node. Here we have used that the flavor central charge of a chiral multiplet in a representation $\bm{R}$ of $G$ is
\begin{equation}
    k^{\text{chiral}}_G = I(\bm{R}) \,,
\end{equation}
where $I(\bm{R})$ is the Dynkin index of $\bm{R}$.

In this paper, we focus on the configurations that may permit identical central charges; this means that we do not consider theories involving the conformal matter building blocks, and the only chiral multiplets that we are allowed to include are adjoint-valued, as per the analysis in \cite{Kang:2021lic}. Such gaugings can only involve a single gauge node and the condition on the $\beta$-function in equation \eqref{eqn:betaone} becomes
\begin{equation}
    \beta_G \leq 0 \quad \Longleftrightarrow \quad \sum_{\mathcal{D}_{p_i}(G)s} \frac{2(p_i - 1)}{p_i}h_G^\vee + \sum_{\text{chirals}} 2h_G^\vee \leq 6h_G^\vee \,.
\end{equation}
It is straightforward to see that there can be at most six $\mathcal{D}_{p_i}(G)$ theories and three adjoint-valued chiral multiplets attached to the $\mathcal{N}=1$ gauge node \cite{Kang:2021lic}. The resulting quivers are all of the form
\begin{equation}
  \begin{gathered}
  \begin{tikzpicture}
    \node[tnode] (N1) {$\mathcal{D}_{p_1}(G)$};
    \node[node,ball color = gray!10, opacity = 0.8] (N2) [right=4mm of N1] {$G$};
    \node[tnode] (N3) [right=4mm of N2] {$\mathcal{D}_{p_6}(G)$};
    \draw (N1.east) -- (N2.west);
    \draw (N2.east) -- (N3.west);
    \draw[->,dashed] (N2) to[out=130, in=410, looseness=8] node[above] {\small $n_a \leq 3$} (N2);
    \draw (N2.216)--++(216:4mm);
    \draw (N2.252)--++(252:4mm);
    \draw (N2.288)--++(288:4mm);
    \draw (N2.324)--++(324:4mm);
    \node[tnode] (T) at (1.6,-1) {$\cdots$};
  \end{tikzpicture}
  \end{gathered} \,,
\end{equation}
where we denote gauge nodes with a background shading as $\mathcal{N}=1$ gauge nodes to differentiate from the $\mathcal{N}=2$ gauge nodes (which are unshaded), and the dashed line indicates $n_a \leq 3$ adjoint chirals. The various combinations of $p_i$ and $n_a$ were listed in \cite{Kang:2021lic}, and we do not repeat them here. 

When the one-loop $\beta$-function for the gauge coupling vanishes (i.e., $\beta_G=0$) then the gauge theory is conformal. If $\beta_G < 0$, however, then we must consider the renormalization group flow to the infrared fixed point of the gauged theory. This fixed point may or may not realize an interacting SCFT. For example, the superconformal R-symmetry, as determined via $a$-maximization \cite{Intriligator:2003jj}, may be inconsistent or the infrared theory may involve a decoupled free sector. In \cite{Kang:2021lic}, the analysis presented therein demonstrates that, if $\gcd(p_i, h_G^\vee) = 1$ for all of the $p_i$ involved in the gauging and there exists an interacting SCFT in the infrared (without introducing new degrees of freedom such as flipper-fields), then the infrared SCFT has identical central charges $a=c$. Interestingly, these conditions appear to be satisfied in most cases where $\beta_G<0$ and one obtains vast families of 4d $\mathcal{N}=1$ SCFTs with $a = c$ \cite{Kang:2021lic}.

\section{Unitarity and superconformal indices}
\label{sec:unitarity}

The superconformal index counts short-multiplets up to recombination into long multiplets. It admits a trace formula as follows: pick a pair of supercharges, $Q$, $Q^\dagger$, and consider the index of the form
\begin{align}
    I(\beta; \mu) = \trace (-1)^F e^{-\beta \delta} e^{-\mu_i \gamma_i} \,, 
\end{align}
where the trace is taken over the Hilbert space of the theory, $F$ is the fermion number, $\delta \equiv \{ Q, Q^\dagger \}$, $\gamma_i$ are generators of the global symmetry algebras of the theory that commute with $Q, Q^\dagger$, and $\mu_i$ are the corresponding chemical potentials. By the usual arguments \cite{Witten:1982df,Kinney:2005ej, Romelsberger:2005eg}, the index only gets contributions from the states satisfying $\delta = 0$, and thus the index is independent of the fugacity $\beta$.

For the case of 4d $\mathcal{N}=1$ superconformal theory, the superconformal algebra is $SU(2, 2|1)$ which has the bosonic subgroup $SO(4, 2) \times U(1)_R$. There are two generators in the superconformal algebra that commute with $Q$, $Q^\dagger$. Choosing $Q=\widetilde{Q}_{\dot{-}}$, we write the index (often called the right-handed index) for a generic $\CN=1$ SCFT as 
\begin{align}
    I(t,y;\bm{v})=\trace(-1)^F t^{3(R+2j_2)}y^{2j_1}\prod_i v_i^{f_i} \,,
\end{align}
where $R$ is the $U(1)$ R-charge, $j_1$ and $j_2$ are the Lorentz spins, and $f_i$ collectively denotes the generators of the flavor symmetries of the theory. The trace is taken over the states with scaling dimension $\Delta$ satisfying 
\begin{align}
    \Delta=\frac{3}{2}R+2j_2 \,. 
\end{align}
The short superconformal multiplets that contribute to this right-handed superconformal index, together with their contribution, are given in Table \ref{tbl:multipletsci}. It is convenient to define the reduced superconformal index which is given by 
\begin{align}\label{eqn:reducedindex}
    \widehat{I}=(1-t^3y)(1-t^3/y)(I-1) \,.
\end{align}
This form is useful since it removes the contributions from the conformal descendants.

Despite the cancellations from the $(-1)^F$ factor, the superconformal index is still powerful enough to determine part of the operator spectrum, in particular for the low-lying operators. We study the spectrum of operators in great detail with the superconformal indices; more specifically, we test the unitarity condition on the structure of the index that operators should satisfy \cite{Beem:2012yn,Evtikhiev:2017heo}. This is important because it is possible for a ``candidate'' superconformal theory to violate unitarity, which is not readily visible at the level of the chiral ring \cite{Maruyoshi:2018nod}.  Furthermore, we investigate various chiral ring operators and observe chiral ring relations, which lift some of the ``classical'' operators.

\begin{table}[H]
    \centering
    \small
    \renewcommand{\arraystretch}{2.0}
    \begin{threeparttable}
    \begin{tabular}{cccc}
    \toprule
    Short mult. & CDI notation & Unitarity condition & Contribution to the index \\\midrule
    $\overline{\mathcal{B}}_{r(j_1,0)}$ & $L\overline{B}_{\overline{1}}[j_1,0]_{\frac{3}{2}r}^{(r)}$ & $r\geq\frac{2}{3}(j_1+1)$ & $(-1)^{2j_1}\dfrac{t^{3r}\chi_{\bf{2j_1+1}}(y)}{(1-t^3y)(1-t^3/y)}$ \\
    $\overline{\mathcal{C}}_{r(j_1,j_2)}$ & $L\overline{A}_{\overline{\ell}}[j_1,j_2]^{(r)}_{2+2j_2+\frac{3}{2}r}$ & $r>\frac{2}{3}(j_1-j_2)$ & $(-1)^{2j_1+2j_2+1}\dfrac{t^{3(r+2j_2+2)}\chi_{\bf{2j_1+1}}(y)}{(1-t^3y)(1-t^3/y)}$\\
    $\widehat{\mathcal{C}}_{(j_1,j_2)}$ & $A_{\ell}\overline{A}_{\overline{\ell}}[j_1,j_2]_{2+j_1+j_2}^{(\frac{2}{3}(j_1-j_2))}$ & -- & $(-1)^{2j_1+2j_2+1}\dfrac{t^{2j_1+4j_2+6}\chi_{\bf{2j_1+1}}(y)}{(1-t^3y)(1-t^3/y)}$\\
    $\mathcal{D}_{(0,j_2)}$ & $B_1\overline{A}_{\overline{\ell}}[0,j_2]_{1+j_2}^{(-\frac{2}{3}(j_2+1))}$ & -- & $-\dfrac{t^{4j_2+4}}{(1-t^3y)(1-t^3/y)}$\\
    $\overline{\mathcal{D}}_{(j_1\geq \frac{1}{2},0)}$ & $A_1\overline{B}_{\overline{1}}[j_1,0]_{1+j_1}^{(\frac{2}{3}(j_1+1))}$ & -- & $(-1)^{2j_1}\dfrac{t^{2j_1+2}\chi_{\bf{2j_1+1}}(y)-t^{2j_1+5}\chi_{\bf{2j_1}}(y)}{(1-t^3y)(1-t^3/y)}$\\
    $\overline{\mathcal{D}}_{(0,0)}$ & $A_2\overline{B}_{\overline{1}}[0,0]_{1}^{(\frac{2}{3})}$ & -- &  $\dfrac{t^{2}}{(1-t^3y)(1-t^3/y)}$\\\bottomrule
    \end{tabular}
    \end{threeparttable}
    \caption{List of $\CN=1$ short multiplets that contribute to the right-handed index, the unitarity conditions they satisfy, and their contributions to the right-handed index. The 4d $\CN=1$ short multiplet contributions to the superconformal index appear in \cite{Gadde:2010en}. Note that the spins $j_1,j_2$ are integer-quantized in \cite{Cordova:2016emh}, whereas they are half-integer-quantized here.}
    \label{tbl:multipletsci}
\end{table}

Here we introduce some terms that, if they were to appear in the superconformal index, would indicate that unitarity of the theory is violated. Any term of the form 
\begin{subequations}
\begin{align}
    t^\lambda\chi_{\bm{2j_1+1}}(y)\quad & (\lambda<2+2j_1) \,,\\ (-1)^{2j_1+1}t^\lambda\chi_{\bm{2j_1+1}}(y)\quad & (2+2j_1\leq\lambda<6+2j_1) \,,
\end{align} 
\end{subequations}
reveals the existence of non-unitary operators \cite{Beem:2012yn,Evtikhiev:2017heo}. Here $\chi_{\bm{2j_1+1}}(y)$ indicates the character of the $\bm{2j_1+1}$ representation of the Lorentz $SU(2)_1$. For small values of $j_1$, we provide the form of the terms which are thus required to be absent below for convenience:
\begin{itemize}
    \item\ $ t^\lambda,\quad\lambda<2$, 
    \item\ $ - t^\lambda,\quad 2 \le \lambda <6$, 
    \item\ $ t^\lambda \chi_{\bm{2}}(y),\quad\lambda<7$, 
    \item\ $ t^\lambda \chi_{\bm{3}}(y),\quad\lambda<4$, 
    \item\ $-t^\lambda \chi_{\bm{3}}(y),\quad 4 \le \lambda <8$.
\end{itemize}

For a sample of the theories with $a=c$ that we construct by gauging $\mathcal{D}_p(G)$ theories, we confirm that they indeed pass this refined unitarity test. In particular, we checked explicitly for those with low rank $G$ by verifying that their superconformal indices do not contain any terms violating unitarity. This goes beyond the study of the chiral operators from \cite{Kang:2021ccs}. 

\section{Superconformal indices of \texorpdfstring{\boldmath{$\CN=2$}}{N=2} \texorpdfstring{\boldmath{$\mathcal{D}_p(G)$}}{Dp(G)} theories}
\label{sec:indexDpG}

To determine the superconformal indices of the $\mathcal{N}=1$ gaugings that we consider in this paper, we first collect the superconformal indices of some individual $\mathcal{D}_p(G)$ theories. 
We define the superconformal index of an $\CN=2$ SCFT as
\begin{align}
    I = \trace (-1)^F t^{2(\Delta + j_2)} y^{2j_1} v^{2R-r} \ ,  
\end{align}
where $R, r$ denote the Cartan generators of the $SU(2)_R \times U(1)_r$ symmetry. The index gets contributions only from the the states satisfying\footnote{Our normalization of $U(1)_r$ is chosen such that the Coulomb branch operators have scaling dimension $\Delta = -r/2$.}
\begin{align}
     \Delta - 2j_2 - 2R - r/2 = 0 .
\end{align}

In particular, we consider $\mathcal{D}_2(SU(3))$, $\mathcal{D}_3(SU(2))$, and $\mathcal{D}_5(SU(2))$ theories.\footnote{These theories are often referred to under different names. In particular, $\mathcal{D}_2(SU(3))=(A_1, D_4)$, $\mathcal{D}_3(SU(2))=(A_1, A_3)=(A_1, D_3)$, and $\mathcal{D}_5(SU(2))=(A_1, D_5)$.} 
The (reduced) $\mathcal{N}=1$ superconformal indices of these theories are given following \cite{Maruyoshi:2016aim, Agarwal:2016pjo}:
\begin{subequations}\label{eqn:Dpindices}
\small
\begin{align}
\begin{split}
    \widehat{I}_{\mathcal{D}_2(SU(3))}=&\ 
    t^3v^{-3}+t^4 \left( v^2\chi_{\mathfrak{su}_3,\bm{8}}-v^{-1}\chi_{\bm{2}}(y)\right)+t^5v+t^6\left(v^{-6}-\chi_{\mathfrak{su}_3,\bm{8}}(z_1,z_2)-1\right)\\
    &+t^7\chi_{\bm{2}}(y)\left(v^2-v^{-4}\right)+t^8\left(2v^{-2}+v^4\chi_{\mathfrak{su}_3,\bm{27}}(z_1,z_2)\right)+t^9\left(v^{-9}-v^{-3}-\chi_{\bm{2}}(y)\right)\\
    &+O\left(t^{10}\right)\,,
\end{split}\\
\begin{split}
    \widehat{I}_{\mathcal{D}_3(SU(2))}=&\ 
    t^{8/3}v^{-8/3}-t^{11/3}v^{-2/3}\chi_{\bm{2}}(y)+t^{4}v^2\chi_{\mathfrak{su}_2,\bm{3}}(z)+t^{14/3}v^{4/3}+t^{16/3}v^{-16/3}\\
    &-t^6\left(\chi_{\mathfrak{su}_2,\bm{3}}(z)+1)\right)-t^{19/3}v^{-10/3}\chi_{\bm{2}}(y)+t^7v^2\chi_{\bm{2}}(y)+t^{22/3}v^{-4/3}\\
    &+t^8\left(v^4\chi_{\mathfrak{su}_2,\bm{5}}(z)+v^{-2}+v^{-8}\right)-t^{26/3}v^{-8/3}-t^9\left(1+v^{-6}\right)+O\left(t^{\frac{28}{3}}\right)\,,
\end{split}\\
\begin{split}
    \widehat{I}_{\mathcal{D}_5(SU(2))}=& \
    t^{12/5}v^{-12/5}+t^{16/5}v^{-16/5}-t^{17/5}v^{-2/5}\chi_{\bm{2}}(y)+t^4v^{2}\chi_{\mathfrak{su}_2,\mathbf{3}}(z)-t^{21/5}v^{-6/5}\chi_{\bm{2}}(y)\\
    &+t^{22/5}v^{8/5}+t^{24/5}v^{-24/5}+t^{26/5}v^{4/5}+t^{28/5}v^{-28/5}-t^{29/5}v^{-14/5}\chi_{\bm{2}}(y)\\
    &-t^6\left(\chi_{\mathfrak{su}_2,\bm{3}}(z)+1\right)+O(t^{32/5})\,,
\end{split}
\end{align}
\end{subequations}
where $\chi_{\bm{2j_1+1}}(y)$ is the character for the $2j_1+1$-dimensional representation of the $SU(2)_1$ factor of the Lorentz group, $\chi_{\mathfrak{su}_N,\bm{R}}(z_1, \cdots,  z_{N-1})$ is the character of the representation $\bm{R}$ of $SU(N)$, $v$ is associated to the $U(1)$ flavor symmetry 
\begin{align}
    \mathcal{F}=-r+2R \,, 
\end{align}
coming from the decomposition of the $\mathcal{N}=2$ R-symmetry, and $z_1, \cdots z_{N-1}$ are the fugacities of the $SU(N)$ flavor symmetry.\footnote{The $U(1)$ flavor symmetries are normalized differently from \cite{Maruyoshi:2016aim}. Our $v$ corresponds to their $v^{-2}$.}

To introduce and explain the concept of operator spectroscopy, such as in \cite{Beem:2012yn,Rastelli:2016tbz}, we first explain how the relevant and marginal operator content of the $\mathcal{D}_p(G)$ SCFTs can be determined from the indices in equation \eqref{eqn:Dpindices}. We can already see some interesting and noteworthy information from these expressions. For example, the moment map operator $\mu$ contributes to the index as $t^4 v^{2}$. We can see that $\mathcal{D}_2(SU(3))$, $\mathcal{D}_3(SU(2))$, and $\mathcal{D}_5(SU(2))$ each have no operator $\trace \mu^2$, as there is no $t^8v^4$ term appearing in either index; these operators are lifted from the spectrum due to a chiral ring relation (also known as the Joseph relation)
\begin{align}
    \mu^2 \Big|_{\mathcal{I}_2} = 0 \,. 
    \label{eq:Joseph}
\end{align}
Here, the Joseph ideal $\mathcal{I}_2$ is defined via 
\begin{align}
    \operatorname{Sym}^2 (\textbf{adj}) = [2 \cdot \textbf{adj}] \oplus \mathcal{I}_2 \,,
    \label{eqn:joseph_ideal}
\end{align}
where $[2 \cdot \textbf{adj}]$ denotes the representation that has highest weight being twice of the Dynkin indices for the adjoint representation. It is well-known that this relation is true for all the $\CN=2$ theories with Higgs branch given as the one-instanton moduli space for $G$, which is identical to the minimal nilpotent orbit of $G$. In fact, for those theories the only non-vanishing part of the $k-$fold product of the moment map operator is in the representation $[k \cdot \textbf{adj}]$, which can be deduced from the universal formula for the Higgs branch Hilbert series \cite{VinbergPopov, Garfinkle, Benvenuti:2010pq, Keller:2011ek, Keller:2012da} or Hall--Littlewood index \cite{Gadde:2011uv, Gaiotto:2012uq} 
\begin{align}
    I_{HL}(\tau) = \sum_{k=0}^{\infty} \chi_{[k \cdot \textbf{adj}]} \tau^{2k} \ ,
\end{align}
where the moment map $\mu$ contributes $\chi_{[\textbf{adj}]} \tau^2$. In particular, the index shows, for all $k$, that 
\begin{align} \label{eq:trmuk}
    \trace \mu^k = 0 \ , 
\end{align}
for the $\mathcal{D}_{p_\text{odd}}(SU(2))$ and $\mathcal{D}_2(SU(3))$ theories since their Higgs branches are the $SU(2)$ and $SU(3)$ (centered) one-instanton moduli spaces, respectively. 
In fact, the relation in equation \eqref{eq:trmuk} holds for all the $\mathcal{D}_p(G)$ theories with $(p, h^\vee)=1$ since their Higgs branches are given by a nilpotent orbit. 

Additional relations among the BPS operators (beyond the Higgs branch operators) in the Argyres--Douglas theories can be deduced \cite{Xie:2021omd, Song:2021dhu}. 
To do this, it is useful to rewrite the superconformal indices of the above $\CN=2$ SCFTs in terms of their short multiplet contents. For the theories we consider, we have 
\begin{align}
\begin{split}
    I_{\mathcal{D}_2 (SU(3))} &= \textrm{PE} \Big[ \chi_{\textbf{adj}} \widehat{\mathcal{B}}_1 + \overline{\mathcal{E}}_{-\frac{3}{2}} + \widehat{\mathcal{C}}_{0(0, 0)} - \widehat{\mathcal{B}}_2 (1 + \chi_{\textbf{adj}}) - \overline{\mathcal{B}}_{1, -2(0, 0)} \\ 
    & \qquad \qquad - \chi_{\textbf{adj}} \overline{\mathcal{B}}_{1, -\frac{3}{2}(0, 0)} - \overline{\mathcal{C}}_{\frac{1}{2}, -1(\frac{1}{2}, 0)} - \widehat{\mathcal{C}}_{1(0, 0)} \chi_{\textbf{adj}} + O(t^{11}) \Big] \ , 
\end{split} \label{eq:D2SU3idxShortMult}\\
\begin{split}
    I_{\mathcal{D}_3 (SU(2))} &= \textrm{PE} \Big[ \chi_{\textbf{adj}} \widehat{\mathcal{B}}_1 + \overline{\mathcal{E}}_{-\frac{4}{3}} +\widehat{\mathcal{C}}_{0(0, 0)} - \widehat{\mathcal{B}}_2 - \overline{\mathcal{B}}_{1, -\frac{5}{3}(0, 0)} \\ 
    & \qquad \qquad - \chi_{\textbf{adj}} \overline{\mathcal{B}}_{1, -\frac{4}{3}(0, 0)} - \overline{\mathcal{C}}_{\frac{1}{2}, -\frac{5}{6}(\frac{1}{2}, 0)} - \widehat{\mathcal{C}}_{1(0, 0)} \chi_{\textbf{adj}} + O(t^{10}) \Big] \ , 
\end{split} \label{eq:D3SU2idxShortMult}\\
\begin{split}
    I_{\mathcal{D}_5 (SU(2))} &= \textrm{PE} \Big[ \chi_{\textbf{adj}} \widehat{\mathcal{B}}_1 + \overline{\mathcal{E}}_{-\frac{6}{5}} + \overline{\mathcal{E}}_{-\frac{8}{5}} +\widehat{\mathcal{C}}_{0(0, 0)} -\overline{\mathcal{B}}_{1, -\frac{8}{5}(0, 0)} \chi_{\textbf{adj}} -\widehat{\mathcal{B}}_2  \\
    & \qquad \qquad - \overline{\mathcal{B}}_{1, -\frac{7}{5}(0, 0)} - \overline{\mathcal{B}}_{1, -\frac{9}{5}(0, 0)} - \overline{\mathcal{B}}_{1, -\frac{11}{5}(0, 0)} + O(t^9) \Big] \ . 
\end{split} \label{eq:D5SU2idxShortMult}
\end{align}
Here, the PE stands for the plethystic exponential and we used various symbols for the short multiplets (in the notation of \cite{Dolan:2002zh}) to denote their indices \cite{Gadde:2011uv}. See Table \ref{tbl:N2multipletsci} for details. In the expressions \eqref{eq:D2SU3idxShortMult} \eqref{eq:D3SU2idxShortMult}, and \eqref{eq:D5SU2idxShortMult}, the multiplets with the positive sign can be thought of as generators and the ones with minus sign as relations. For example, the $\widehat{\mathcal{B}}_1$ is a conserved current multiplet for the $SU(2)$ or $SU(3)$ flavor symmetry. We see there is a term $\widehat{\mathcal{B}}_2$ with negative sign, which translates to the Joseph relation, in equation \eqref{eq:Joseph}, for the Higgs branch operators. The index also allows us to study relations beyond the Schur sector, as discussed in \cite{Song:2021dhu, Xie:2021omd}. See also \cite{Bhargava:2022cuf} for a detailed study on the operator spectrum of minimal Argyres--Douglas theory. 

The $\overline{\mathcal{E}}_r$ multiplet contains the Coulomb branch operator $u$ with dimension $\Delta = -r$, which is the top component in this multiplet.\footnote{As we are considering the right-handed index, we are sensitive to the anti-holomorphic $\overline{\mathcal{E}}$-multiplets; equation \eqref{eqn:EtoB} describes the decomposition of the $\overline{\mathcal{E}}$-multiplet, however we will generally abuse notation and write the $\overline{\mathcal{E}}$, and its superconformal primary, without the bars.} In terms of $\CN=1$ multiplets, it decomposes as
\begin{align}\label{eqn:EtoB}
    \overline{\mathcal{E}}_{-r} \rightarrow \overline{\mathcal{B}}_{\frac{r}{3}(0, 0)} \oplus \overline{\mathcal{B}}_{\frac{r+1}{3}(0, \frac{1}{2})} \oplus \overline{\mathcal{B}}_{\frac{r+2}{3}(0, 0)} \, . 
\end{align}
The $\mathcal{N}=1$ superconformal primary of the first multiplet is $u$, and that of the next two are $Qu$ and $Q^2u$, respectively, where $Q$ is the $\mathcal{N}=2$ supercharge which is not the $\mathcal{N}=1$ supercharge.

\begin{table}[H]
    \centering
    \small
    \renewcommand{\arraystretch}{2.0}
    \begin{threeparttable}
    \begin{tabular}{ccc}
    \toprule
    Short mult. & CDI notation & Contribution to the index \\\midrule
    $\widehat{\mathcal{B}}_R$ & $B_1 \overline{B}_{\overline{1}}[0;0]^{(R; 0)}$ & $\dfrac{-t^{2+4R} v^{2(R-1)} + t^{4R}v^{2R}}{(1-t^3 y)(1-t^3/y)}$\\  
    $\overline{\mathcal{B}}_{1, r(0, 0)}$ & $L \overline{B}_{\overline{1}}[0;0]^{(1, -r)}$ & $ - \dfrac{t^{4-2r} v^{2r}(t^2 - v^2) (1+t^2 v^4 - t v^2 \chi_{\bf{2}}(y))}{(1-t^3 y)(1-t^3/y)}$ \\  
    $\widehat{\mathcal{C}}_{R(0, 0)}$ & $A_{\ell} \overline{A}_{\overline{\ell}}[0; 0]^{(R;0)}$ & $\frac{ t^{6+4R}v^{2R-2} (t^2 - v^2) (1 -tv^2 \chi_{\bf{2}}(y)) }{(1-t^3 y)(1-t^3/y)}$ \\
    $\overline{\mathcal{C}}_{R, r(j_1, 0)}$ & $L \overline{A}_{\overline{\ell}} [j_1; 0]^{(R;-r)}$ & $\dfrac{(-1)^{2j_1} t^{6-2r+4R} v^{2(R+r-1)}(t^2-v^2)(1+t^2 v^2 - t v^2 \chi_{\bf{2}}(y) )\chi_{\bf{2j_1}}(y) }{(1-t^3 y)(1-t^3/y)}  $ \\
    $\overline{\mathcal{E}}_{r}$ & $ L \overline{B}_{\overline{1}} [0;0]^{(0;-r)}$ & $\dfrac{t^{-2r}v^{2r}(1+t^2 v^4 - tv^2 \chi_{\bf{2}}(y))}{(1-t^3 y)(1-t^3/y)}$
    \\ \bottomrule
    \end{tabular}
    \end{threeparttable}
    \caption{List of $\CN=2$ short multiplets that appear in the indices in equations \eqref{eq:D2SU3idxShortMult}, \eqref{eq:D3SU2idxShortMult}, and \eqref{eq:D5SU2idxShortMult}, and their contributions to the index \cite{Gadde:2011uv}. We denote the charge from the $U(1)$ R-symmetry as $r$ and the charge from the $SU(2)$ R-symmetry as $R$.}
    \label{tbl:N2multipletsci}
\end{table}

\section{\texorpdfstring{\boldmath{$\CN=2$}}{N=2} SCFTs: \texorpdfstring{\boldmath{$\widehat{D}_4(SU(3))$}}{D4hat(SU3)} and \texorpdfstring{\boldmath{$\widehat{E}_6(SU(2))$}}{E6hat(SU2)} theories}
\label{sec:indexD4E6}

Before studying the spectrum of various $\mathcal{N}=1$ superconformal field theories with $a=c$, we first look into the spectrum of $\mathcal{N}=2$ SCFTs of type $\widehat{\Gamma}(G)$ considered in \cite{Kang:2021lic}. With the building blocks in hand, we study the $\widehat{D}_4(SU(3))$ theory and the $\widehat{E}_6(SU(2))$ theory. The $\widehat{D}_4(SU(3))$ theory can be constructed via gauging four copies of $\mathcal{D}_2(SU(3))$, whereas the $\widehat{E}_6(SU(2))$ theory can be constructed with three copies of $\mathcal{D}_3(SU(2))$:
\begin{subequations}
\begin{align}
\widehat{D}_4(SU(3))\ :\quad &
\begin{aligned}
    \begin{tikzpicture}
      \node[node] (s0) {$SU(3)$};
      \node[d2] (c1) [left=0.6cm of s0] {$\mathcal{D}_{2}(SU(3))$};
      \node[d2] (c2) [right=0.6cm of s0] {$\mathcal{D}_{2}(SU(3))$};
      \node[d2] (c3) [above=0.4cm of s0] {$\mathcal{D}_{2}(SU(3))$};
      \node[d2] (c4) [below=0.4cm of s0] {$\mathcal{D}_{2}(SU(3))$};
      \draw (s0.east) -- (c2.west);
      \draw (s0.west) -- (c1.east);
      \draw (s0.north) -- (c3.south);
      \draw (s0.south) -- (c4.north);
    \end{tikzpicture} 
\end{aligned}\,,\\
\widehat{E}_6(SU(2))\ :\quad &
\begin{aligned}
    \begin{tikzpicture}
      \node[node] (s0) {$SU(2)$};
      \node[d2] (c1) [left=0.6cm of s0] {$\mathcal{D}_{3}(SU(2))$};
      \node[d2] (c2) [right=0.6cm of s0] {$\mathcal{D}_{3}(SU(2))$};
      \node[d2] (c3) [above=0.4cm of s0] {$\mathcal{D}_{3}(SU(2))$};
      \draw (s0.east) -- (c2.west);
      \draw (s0.west) -- (c1.east);
      \draw (s0.north) -- (c3.south);
    \end{tikzpicture} 
\end{aligned}\,.
\end{align}
\end{subequations}
The superconformal indices are computed utilizing the expression
\begin{align}\label{eqn:frenchtoast}
\begin{split}
    I_{\mathcal{N}=2}=\frac{1}{|W_G|}\oint[dz]_G\ I^G_{\textrm{vec}}(z) \prod_{i=1}^n I_{\mathcal{D}_{p_i}(G)}(t,y;z,v) , 
\end{split}
\end{align}
where $I^G_{\textrm{vec}}(z)$ denotes the index for the vector multiplet (for gauge group $G$) given as
\begin{align}
    I^G_{\textrm{vec}}(z) = \operatorname{PE}\left[\frac{-t^3y-t^3/y+2t^6+t^2v^{-2}-t^4v^2}{(1-t^3y)(1-t^3/y)}\, \chi_{G,\textbf{adj}}(z)\right] \,,
\end{align}
and where $[dz]_G$ is the integration measure for $G$:
\begin{align} \label{eq:Haar}
    [dz]_G = \prod_{i=1}^r \frac{dz_i}{2\pi iz_i} \prod_{\alpha\in\text{Roots}}(1-z^{\bm{\alpha}}) \,.
\end{align}
Here, $|W_G|$ is the dimension of Weyl group of $G$, $r$ is the rank of $G$, and the product is over all roots of $G$. Furthermore, $\bm{z}$ denotes the fugacities of $G$, and PE is the plethystic exponential.

\subsection{\texorpdfstring{$\widehat{D}_4(SU(3))$}{D4hat(SU3)} theory}

Applying the formula in equation \eqref{eqn:frenchtoast}, we find that the reduced index of $\widehat{D}_4(SU(3))$ is
\begin{align}
    \begin{split}
        \widehat{I}_{\widehat{D}_4(SU(3))}=&\ 4t^3v^{-3}+t^4\left(v^{-4}-4v^{-1}\chi_{\bm{2}}(y)\right)+t^5\left(4v-v^{-2}\chi_{\bm{2}}(y)\right)+11t^6v^{-6}\\
        &+t^7\left(4v^{-7}+\chi_{\bm{2}}(y)\left(v^2-17v^{-4}\right)\right)\\
        &+t^8\left(v^{-8}+27v^{-2}+2v^4-8v^{-5}\chi_{\bm{2}}(y)+6v^{-2}\chi_{\bm{3}}(y)\right)\\
        &+t^9\left(4v^{-3}+24v^{-9}+\chi_{\bm{2}}(y)\left(-16+5v^{-6}\right)+4v^{-3}\chi_{\bm{3}}(y)\right)+O(t^{10}).
    \end{split}
\end{align}
Let us explain some of the operator spectrum that we observe from the index:
\begin{itemize}
    \item $4 t^3 v^{-3}$: This term comes from the Coulomb branch operators (i.e., the $\CN=2$ chiral operators) $u_i$ of dimension $3/2$ in each of the four $\mathcal{D}_2(SU(3))$ theories. 
    \item $t^4v^{-4}$: It is associated to the operator $\trace \phi^2$, where $\phi$ is the adjoint chiral in the $\CN=2$ vector multiplet. 
    \item $4 t^5 v^{1}$: It arises from the $\CN=2$ super-descendants of the Coulomb branch operators in the first bullet point. These are $Q^2 u_i$, where $Q$ is the $\mathcal{N}=2$ supercharge of each $\mathcal{D}_2(SU(3))$ theory with non-zero $j_2$ and non-trivial $\CN=2$ $U(1)$ R-charge.
    \item $11 t^6 v^{-6}$: This term comes from ten marginal operators of the form $u_i u_j$ and another one given by $\trace \phi^3$. The operators of the form $\trace \phi \mu_i$ are absent due to the F-term relation and the chiral ring relation in equation \eqref{eq:trmuk}.
 \end{itemize}
All relevant and marginal operators, including those with non-zero $j_1$, are listed in Table \ref{tbl:2s}.

For an $\mathcal{N}=2$ theory, we can take the Macdonald limit \cite{Gadde:2011uv}, which is defined as $(t^3y)\rightarrow 0$ while $(t^3/y)=q$ and $T=t v^2y$ are held fixed. Then, the Macdonald index is given as
\begin{align}
    I^{\text{Mac}}=\trace (-1)^F q^{\Delta-R}T^{R-r/2} \,,
\end{align}
where the trace is taken over the operators with $\Delta-2j_2-2R-r/2=0$ and $j_1+j_2+r/2=0$. We compute the Macdonald index of the $\widehat{D}_4(SU(3))$ theory to be
\begin{align}
\begin{split}
    I^{\text{Mac}}_{\widehat{D}_4(SU(3))}=&\ 1+(T+2T^2)q^2+(T-T^3)q^3+(T+T^2+2T^4)q^4\\&+(T+T^2-T^4-T^5)q^5
    +(T+2T^2+2T^3+2T^6)q^6\\&+(T+2T^2+2T^3-2T^4-2T^5-T^7)q^7\\
    &+(T+3T^2+3T^3-T^4-2T^5+2T^8)q^8\\&+(T+3T^2+4t^3-2T^4-4T^5-T^6-T^9)q^9
    +O\left(q^{10}\right) \,.
\end{split}
\end{align}
We further take the Schur limit $(T\rightarrow 1)$ from the Macdonald index to obtain the Schur index, which is defined as
\begin{align}
    I^{\text{Schur}}=\trace\,(-1)^Fq^{\Delta-R} \,.
\end{align}
We compute the Schur index of $\widehat{D}_4(SU(3))$ as
\begin{align}
I^{\text{Schur}}_{\widehat{D}_4(SU(3))}=1+3q^2+4q^4+7q^6+6q^8+O\left(q^{10}\right)\,,   
\end{align} 
which matches with the known result \cite{Kang:2021lic}. The Schur sector of any $\CN=2$ SCFT is captured by its vertex operator algebra (VOA) \cite{Beem:2013sza}. It would be interesting to directly construct the VOA for the $\widehat{\Gamma}(G)$ theories to gain further access on this theory. 

\subsection{\texorpdfstring{$\widehat{E}_6(SU(2))$}{E6hat(SU2)} theory}

We compute the reduced index of $\widehat{E}_6(SU(2))$ theory and find it as
\begin{align}\label{eqn:E6hatSU2index}
    \begin{split}
        \widehat{I}_{\widehat{E}_6(SU(2))}=&\ 3t^{\frac{8}{3}}v^{-\frac{8}{3}}-3t^{\frac{11}{3}}v^{-\frac{2}{3}}\chi_{\bm{2}}(y)+t^4v^{-4}+3t^{\frac{14}{3}}v^{\frac{4}{3}}-t^5v^{-2}\chi_{\bm{2}}(y)+6t^{\frac{16}{3}}v^{-\frac{16}{3}}\\
        &-9t^{\frac{19}{3}}v^{-\frac{10}{3}}\chi_{\bm{2}}(y)+t^{\frac{20}{3}}\left(3v^{-\frac{20}{3}}+3v^{-\frac{2}{3}}\chi_{\bm{3}}(y)\right)+t^7v^2\chi_{\bm{2}}(y)\\
        &+t^{\frac{22}{3}}\left(12v^{-\frac{4}{3}}+3v^{-\frac{4}{3}}\chi_{\bm{3}}(y)\right)-6t^{\frac{23}{3}}v^{-\frac{14}{3}}\chi_{\bm{2}}(y)+t^8\left(v^{-2}+11v^{-8}\right)\\
        &+t^{\frac{25}{3}}\left(3v^{-\frac{16}{3}}-6v^{\frac{2}{3}}\right)\chi_{\bm{2}}(y)+3t^{\frac{26}{3}}v^{-\frac{8}{3}}\chi_{\bm{3}}(y)-t^9\left(1+19v^{-6}\right)\chi_{\bm{2}}(y)\\
        &+O(t^{10}).        
    \end{split}
\end{align}
Let us explain some of the important operators that can be read off from the index: 
\begin{itemize}
    \item $3t^{8/3}v^{-8/3}$: It arises from the Coulomb branch operator of dimension $3/2$ in each $\CN=2$ $\mathcal{D}_3(SU(2))$ theory, $u_i$.
    \item $t^4 v^{-4}$: It comes from $\trace \phi^2$ where $\phi$ is the scalar in the $\CN=2$ vector multiplet.  
    \item $3 t^{14/3} v^{4/3}$: Super-partners of the Coulomb branch operators of each $\mathcal{D}_2(SU(3))$: $Q^2u_i$. 
    \item $6 t^{16/3} v^{-16/3}$: This term comes from products of Coulomb branch operators: $u_i u_j$. 
\end{itemize}
All of the relevant and marginal operators, including those with a non-zero $j_1$, are listed in Table \ref{tbl:2s}. We can also take the Macdonald limit of equation \eqref{eqn:E6hatSU2index} to get the Macdonald index of $\widehat{E}_6(SU(2))$ theory, which is
\begin{align}
    \begin{split}
        I^{\text{Mac}}_{\widehat{E}_6(SU(2))}=&\
        1+Tq^2+Tq^3+(T-T^3)q^4+(T-T^3)q^5+(T+T^2)q^6\\
        &+(T+T^2-T^3-T^4)q^7+(T+2T^2-T^3-T^4)q^8+O\left(q^9\right) \,.
    \end{split}
\end{align}
We can further take the Schur limit to obtain
\begin{align}
  I^{\text{Schur}}_{\widehat{E}_6(SU(2))}=1+q^2+q^3+2q^6+q^8+O\left(q^9\right)\,.
\end{align}
This result matches with the known result \cite{Buican:2020moo,Kang:2021lic}. The associated vertex operator algebra (VOA) for this theory is given by the $\mathcal{A}(6)$ algebra of \cite{Feigin:2007sp} and studied in detail in \cite{Buican:2016arp, Buican:2020moo}. It would be interesting to reproduce the Macdonald index we presented above from the VOA as was done in other AD theories \cite{Song:2016yfd, Agarwal:2018zqi, Watanabe:2019ssf, Xie:2019zlb}. 

We can also consider the Hall--Littlewood limit of the superconformal index and thus determine the operators which exist in that sector (quite often identical to the set of Higgs branch operators) of the spectrum of the theory; for $a=c$ theories (and beyond), this limit has been investigated thoroughly in \cite{Kang:2022zsl}. The Hall--Littlewood indices for $\widehat{D}_4(SU(2N+1))$ theories, which can be realized as genus-zero class $\mathcal{S}$ theories with $\mathbb{Z}_2$ twist lines, demonstrated the presence of operators belonging to $\mathcal{D}$-type multiplets. This demonstrated that, in contrast to a common belief, the Higgs branch Hilbert series is different from the Hall--Littlewood index, even for genus-zero class $\mathcal{S}$ theories. Among the $\widehat{\Gamma}(G)$ theories, the Hall--Littlewood indices indeed agrees with Higgs branch Hilbert series when $\Gamma = E_6, E_7, E_8$; however, when $\Gamma = D_4$, they do not agree. 

\section{Conformal manifolds and conformal gaugings}
\label{sec:conformal}

The superconformal index allows us to use the technique of operator spectroscopy to determine the low-scaling-dimension operator content of the theories that we consider. These SCFTs are constructed by gauging together a collection of $\mathcal{N}=2$ $\mathcal{D}_p(G)$ theories via an $\mathcal{N}=1$ or $\mathcal{N}=2$ vector multiplet. Understanding the operator content of the individual $\mathcal{D}_p(G)$  theories provides a strong start: this we do by studying the reduced superconformal indices in Section \ref{sec:indexDpG}. However, upon gauging there are many subtleties to take into account; the R-charges of the operators from each $\mathcal{D}_p(G)$ theory are shifted through $a$-maximization, and non-trivial chiral ring relations amongst the operators of the gauged theory can rule out naively expected operators. For this reason, the superconformal index is a vital tool to determine both that the gauged theories are unitary and to extract the operator spectrum.

However, in this section, we consider a slightly simpler class of theories obtained via $\mathcal{N}=1$ gauging of a collection of $\mathcal{D}_{p_i}(G)$ such that the one-loop $\beta$-function of the introduced gauge coupling vanishes. Such $p_i$ were enumerated in \cite{Kang:2021ccs}. For these theories, we now explore the exactly marginal operators, and thus the structure of the $\mathcal{N}=1$ conformal manifold, that can be determined via the study of the spectrum of Coulomb branch scaling dimensions of each individual $\mathcal{D}_{p_i}(G)$. When doing $\mathcal{N}=1$ conformal gauging, it is important to verify that there exists at least one exactly marginal operator, otherwise the gauged theory does not, in fact, give rise to an interacting SCFT \cite{Leigh:1995ep,Green:2010da}. These can be determined independently of the superconformal index, and we verify that the superconformal index for the single $\mathcal{N}=1$ conformal gauging that we determine (see Section \ref{sec:6D23}) matches the counting done in this section. We focus on theories with $a = c$, and thus it is necessary that $\gcd(p_i, h_G^\vee) = 1$ for each $p_i$. Analysis of all of the possible conformal gaugings in \cite{Kang:2021ccs} leads us to conclude that $G = SU(N)$ is a necessary condition for $a = c$. There are eighteen different conformal gaugings, and they contain the following possibilities for $p$:
\begin{equation}
    p \in \{ 2, 3, 4, 5, 6, 7, 8, 9, 10, 12, 15, 18, 20, 24, 42 \} \,.
\end{equation}
We refer to \cite{Kang:2021ccs} for the complete list of combinations.
To determine which products of Coulomb branch operators are marginal, we actually only need to know the Coulomb branch operators which have $\Delta < 2$. If $\mathcal{D}_p(SU(N))$ is such that $N > p$, then the spectrum of Coulomb branch operators, with $\Delta < 2$, contains one operator with each of the following scaling dimensions:
\begin{equation}
    \Delta_{\leq 2} = \bigg\{ \frac{p+1}{p}, \frac{p+2}{p}, \cdots, \frac{2p - 1}{p} \bigg\} \,,
\end{equation}
as can be observed from the general formula for the Coulomb branch scaling dimensions of $\mathcal{D}_p(G)$:
\begin{align}
    \mathcal{C}(p, G) = \left\{j - \frac{h_G^\vee}{p}s \,\Big|\, j - \frac{h_G^\vee}{p}s > 1\,,\, j \in \operatorname{Cas}(G)\,,\, s = 1, \cdots, p-1 \right\} \,.
\end{align}
Here, $\operatorname{Cas}(G)$ are the degrees of the fundamental Casimir invariants of $G$. 

The number of marginal operators formed by the product of two Coulomb branch operators
\begin{equation}
    u_a u_b \,,
\end{equation}
of a $\mathcal{D}_p(N > p)$ theory is given by
\begin{equation}
    \operatorname{\#marginal}(\mathcal{D}_p(SU(N > p))) = \bigg\lceil \frac{p-1}{2} \bigg\rceil \,.
\end{equation}
Next, we consider marginal operators formed from the product of two different Coulomb branch operators belonging to two different Argyres--Douglas theories: $\mathcal{D}_{p_1}(SU(N))$ and $\mathcal{D}_{p_2}(SU(N))$. We consider $N > p_2 \geq p_1$, and let $\ell = \gcd(p_1, p_2)$ denote the greatest common divisor. It is straightforward to see that the number of such marginal operators is
\begin{equation}
    \ell - 1 \,.
\end{equation}
Combining these two results we can determine the number of marginal operators, formed from the product of two Coulomb branch operators of the $\mathcal{D}_p(G)$ building blocks, of the $\mathcal{N}=1$ gauged theory. We find
\begin{equation}\label{eqn:marginal}
    \operatorname{\#\, marginal\ operators} \quad = \quad \sum_{p_i, p_{j>i}} (\gcd(p_i, p_j) - 1) + \sum_{p_i}\bigg\lceil \frac{p_i-1}{2} \bigg\rceil \,,
\end{equation}
where we assume that the $p_i$ are increasing $p_1 \leq \cdots \leq p_6$ and that the gauge group is $SU(N > p_6)$. Furthermore, the total number of $U(1)$ flavor currents in the gauged theory is the number of $p_i > 1$ minus one, and thus we can see from equation \eqref{eqn:marginal} that the number of exactly marginal operators in the theory after the conformal gauging is non-zero. 

It remains for us to consider the number of marginal operators that exist when $N < p$. There is a small finite number of such cases and in each the number of marginal operators on the Coulomb branch can be determined exhaustively. We have verified that, in all cases, the number of exactly marginal operators is strictly positive. Thus, all conformal $(\mathcal{N}=1)$-gaugings of the common $G$ flavor symmetry of a collection of $\mathcal{D}_{p_i}(G)$ theories gives rise to an interacting $\mathcal{N}=1$ SCFT.

\section{\texorpdfstring{\boldmath{$\mathcal{N}=1$}}{N=1} gaugings of multiple \texorpdfstring{\boldmath{$\mathcal{D}_2(SU(3))$}}{D2(SU3)}}
\label{sec:multipleD2SU3}

Next, we turn to the analysis of the superconformal index for strictly $\mathcal{N}=1$ theories. We begin by considering the four theories that are built out of between three and six copies of the $\mathcal{D}_2(SU(3))$ theory via $\mathcal{N}=1$ gauging. The quiver diagrams of these theories are
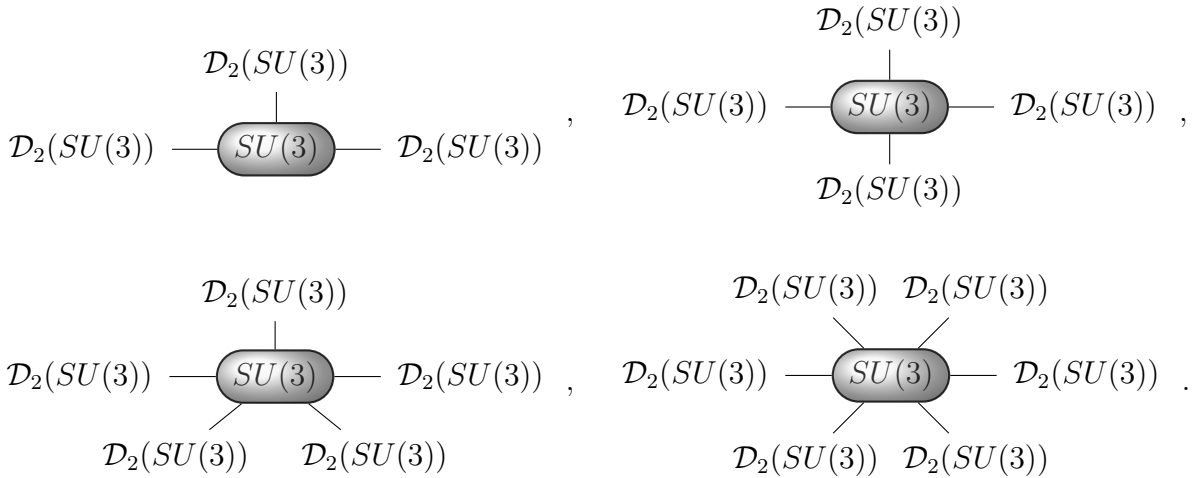
\begin{align*}
\centering
\begin{aligned}
    \begin{tikzpicture}
      \node[gaugeN1,ball color = gray!10, opacity = 0.8] (s0) {$SU(3)$};
      \node[d2] (c1) [left=0.6cm of s0] {$\mathcal{D}_{2}(SU(3))$};
      \node[d2] (c2) [right=0.6cm of s0] {$\mathcal{D}_{2}(SU(3))$};
      \node[d2] (c3) [above=0.4cm of s0] {$\mathcal{D}_{2}(SU(3))$};
      \draw (s0.east) -- (c2.west);
      \draw (s0.west) -- (c1.east);
      \draw (s0.north) -- (c3.south);
    \end{tikzpicture} 
\end{aligned}\,,\quad &
\begin{aligned}
    \begin{tikzpicture}
      \node[gaugeN1,ball color = gray!10, opacity = 0.8] (s0) {$SU(3)$};
      \node[d2] (c1) [left=0.6cm of s0] {$\mathcal{D}_{2}(SU(3))$};
      \node[d2] (c2) [right=0.6cm of s0] {$\mathcal{D}_{2}(SU(3))$};
      \node[d2] (c3) [above=0.4cm of s0] {$\mathcal{D}_{2}(SU(3))$};
      \node[d2] (c4) [below=0.4cm of s0] {$\mathcal{D}_{2}(SU(3))$};
      \draw (s0.east) -- (c2.west);
      \draw (s0.west) -- (c1.east);
      \draw (s0.north) -- (c3.south);
      \draw (s0.south) -- (c4.north);
    \end{tikzpicture} 
\end{aligned}\,,\\[8pt]
\begin{aligned}
    \begin{tikzpicture}
      \node[gaugeN1,ball color = gray!10, opacity = 0.8] (s0) at (0,0) {$SU(3)$};
      \node[d2] (c1) at (90:1.1) {$\mathcal{D}_{2}(SU(3))$};
      \node[d2] (c2) at (180:2.6) {$\mathcal{D}_{2}(SU(3))$};
      \node[d2] (c3) at (219:1.7) {$\mathcal{D}_{2}(SU(3))$};
      \node[d2] (c4) at (321:1.7) {$\mathcal{D}_{2}(SU(3))$};
      \node[d2] (c5) at (0:2.6) {$\mathcal{D}_{2}(SU(3))$};
      \draw (c1)--(s0);
      \draw (c2)--(s0);
      \draw (c3)--(s0);
      \draw (c4)--(s0);
      \draw (c5)--(s0);
    \end{tikzpicture}
\end{aligned}\,,\quad &
\begin{aligned}
    \begin{tikzpicture}
      \node[gaugeN1,ball color = gray!10, opacity = 0.8] (s0) at (0,0) {$SU(3)$};
      \node[d2] (c1) at (0:2.6) {$\mathcal{D}_{2}(SU(3))$};
      \node[d2] (c2) at (45:1.6) {$\mathcal{D}_{2}(SU(3))$};
      \node[d2] (c3) at (135:1.6) {$\mathcal{D}_{2}(SU(3))$};
      \node[d2] (c4) at (180:2.6) {$\mathcal{D}_{2}(SU(3))$};
      \node[d2] (c5) at (225:1.6) {$\mathcal{D}_{2}(SU(3))$};
      \node[d2] (c6) at (315:1.6) {$\mathcal{D}_{2}(SU(3))$};
      \draw (c1)--(s0);
      \draw (c2)--(s0);
      \draw (c3)--(s0);
      \draw (c4)--(s0);
      \draw (c5)--(s0);
      \draw (c6)--(s0);
    \end{tikzpicture}
\end{aligned}\,.
\end{align*}
The index of the theory arising from gauging of $n$ copies of the $\mathcal{D}_{p_i}(G)$ theory via an $\mathcal{N}=1$ vector multiplet is given by the integral
\begin{align}\label{eqn:swipernoswiping}
   I=\frac{1}{|W_G|} \oint [dz]_G I^G_{\textrm{vec}}(\bm{z}) \prod_{i=1}^nI_{\mathcal{D}_{p_i}(G)}(t,y;\bm{z},v_i t^{3\epsilon_i}) \,,
\end{align}
where $\bm{z}$ denotes fugacities of the introduced gauge group $G$, $\bm{v}$ denotes those of the flavor symmetries, $|W_G|$ is the dimension of Weyl group of $G$, $[dz]_G$ is the integration measure (defined in equation \eqref{eq:Haar}) 
and $I_\textrm{vec}^G$ is the index for the $\CN=1$ vector multiplet with gauge group $G$ which is given by
\begin{align}
    I_{\textrm{vec}}^G (\bm{z}) = \operatorname{PE}\left[\frac{-t^3y-t^3/y+2t^6}{(1-t^3y)(1-t^3/y)}\chi_{G,\textbf{adj}}(\bm{z})\right] \ ,
\end{align}
and PE is the plethystic exponential.

The $v_it^{3\epsilon_i}$ term in the indices of the $\mathcal{D}_{p_i}(G)$ theories arises from the mixing between the UV R-symmetry and Abelian flavor symmetries which forms the superconformal R-charge:
\begin{align}
    U(1)_R^{\text{IR}}=U(1)_R+\epsilon_i\mathcal{F}_i \,.
\end{align}
In addition, there always exists one particular linear combination of the $\mathcal{F}_i$ that is anomalous. It follows that all $v_i$ shall satisfy a corresponding relation; in turn, $v_i$ can be redefined into fugacities of the anomaly-free flavor symmetries. For example, when considering three copies of $\mathcal{D}_2(SU(3))$ gauged together, the diagonal $U(1)$ generated by $\mathcal{F}_1+\mathcal{F}_2+\mathcal{F}_3$ is anomalous and only the axial $U(1)$s generated by $\mathcal{F}_1-\mathcal{F}_2$ and $\mathcal{F}_2 - \mathcal{F}_3$ remain as non-anomalous symmetries of the gauged theory. The diagonal $U(1)$ anomaly imposes the condition that
\begin{align}
    \prod_i v_i=1 \,,
\end{align}
and the index recombines into fugacities of the two axial $U(1)$s,
\begin{align}\label{eqn:tildev}
    \widetilde{v}_i=v_i^{-1}v_{i+1}\,.
\end{align}
In this section we typically turn off the fugacities of the $U(1)$s, i.e., 
\begin{align}
    \widetilde{v}_i = 1, 
\end{align}
as the way in which they enter the index is not of key importance for the purposes of determining unitarity or the spectrum of relevant/marginal operators. As usual, we refer to the version of the index obtained in this way as the \emph{unrefined index}. In Appendix \ref{app:indicesplusplus}, we list the full, refined, expressions for completeness.

\subsection{Gauging 3 copies of \texorpdfstring{$\mathcal{D}_2(SU(3))$}{D2(SU3)}}\label{sec:3D23}

We first compute the reduced index of the theory with three copies of $\mathcal{D}_2(SU(3))$ glued together via $\mathcal{N}=1$ gauging. The coefficients $\epsilon_i$ of the flavor mixing to the infrared R-charges are determined solely by the anomaly-free condition as:
\begin{align}
    \epsilon_i=-\frac{1}{3} \,,\quad i=1,2,3 \,,
\end{align} 
where we note that no operator crosses the unitarity bound along the flow into the infrared. From the mixing parameters, we can determine the R-charge of the various relevant operators after the gauging. We find
\begin{equation}
    R(\mu_i) = \frac{2}{3} \,, \quad R(u_i) = 2 \,, \quad R(Q_i) = -\frac{1}{3} \,,
\end{equation}
for the R-charges of the moment maps, the Coulomb branch operators, and the leftover $\mathcal{N}=2$ supercharges. Computing the reduced unrefined index from the integral in equation \eqref{eqn:swipernoswiping}, we find 
\begin{align}\label{eqn:ind222}
    \widehat{I}_{(2,2,2)}=6t^4-3t^5\chi_{\bm{2}}(y)+3t^6-3t^7\chi_{\bm{2}}(y)+12t^8-11t^9\chi_{\bm{2}}(y)+O(t^{10}) \,.
\end{align}
The first term $6t^4$ comes from the operators of the form $\trace \mu_i \mu_{j \neq i}$ and the $Q^2 u_i$. The $u_i$ where $i=1, 2, 3$ denotes the single Coulomb branch operator in each of the $\mathcal{D}_2(SU(3))$ building blocks, each with scaling dimension $\Delta=\frac{3}{2}$; $Q^2 u_i$ is a superconformal descendant in the $\mathcal{N}=2$ theory, however the $\mathcal{E}$-multiplet decomposes as in equation \eqref{eqn:EtoB} when the gauging breaks the symmetry to $\mathcal{N}=1$, and $Q^2u_i$ corresponds to the primary of an $\mathcal{N}=1$ multiplet. The three marginal operators that contribute to the $t^6$ term correspond to the Coulomb branch operators, $u_i$. In fact, there are two more marginal operators 
\begin{align}
    \trace \mu_1\mu_2\mu_3 \,,\quad \trace \mu_3\mu_2\mu_1 \,,
\end{align}
that are neutral under both Abelian flavor symmetries $\mathcal{F}_1-\mathcal{F}_2$ and $\mathcal{F}_2-\mathcal{F}_3$. The contributions from these operators are canceled precisely by the negative contributions from the two $U(1)$ current multiplets. We do not find the other operators of the form $\trace \mu_i\mu_j\mu_k$ as they are lifted by a chiral ring relation between the adjoint part of $\mu^2$ in the $\mathcal{D}_2(SU(3))$ theories, as we discuss around equation \eqref{eq:Joseph}. There are some other operators which we might expect to contribute to the index in equation \eqref{eqn:ind222}, and for which we now discuss the expected reason for their absence. There are three operators involving the moment maps and the gaugino: $\trace \lambda \mu_i$. We would expect these operators to contribute to the index as
\begin{equation}\label{eqn:doratheexplorer}
 -t^5(v_1^2+v_1^{-2}v_2^2+v_2^{-2})\chi_{\bm{2}}(y) \,,
\end{equation}
where we have restored the $U(1)^2$ flavor symmetry fugacities.\footnote{We emphasize that these are the fugacities associated to the non-anomalous axial $U(1)$ symmetries --  for convenience, we have slightly abused notation by dropping  the tildes compared to equation \eqref{eqn:tildev}.} However, we can see that the superconformal index of the $\mathcal{D}_2(SU(3))$ theory, which is written in equation \eqref{eqn:Dpindices}, has a term $t^7\chi_{\bm{2}}(y)v^2$, which, after gauging, contributes to the $t^5\chi_{\bm{2}}(y)$ term as in equation \eqref{eqn:doratheexplorer}, where the overall coefficient is instead $+1$. We expect that these operators pair up to form long multiplets after gauging, and thus they no longer contribute to the index, which is only sensitive to (certain) short multiplets, as discussed. We note that it is necessary to go to the flavor-fugacity-refined index to see that it is the $\trace \lambda \mu_i$ operators that recombine into long multiplets, and not the $Qu_i$ operators.

To determine the superconformal multiplets that generate the ring of short multiplets, and the relations among them in this ring, it is often helpful to take the plethystic logarithm of the superconformal index. We again reduce by multiplying by $(1-t^3y)(1-t^3/y)$ to remove the contribution from conformal descendants, which renders the generators and ring relations more clearly. We define the reduced plethystic log of the superconformal index as
\begin{align}\label{eqn:plogreduced}
    \widetilde{I}\equiv (1-t^3y)(1-t^3/y)\operatorname{PLog}[I]\,.
\end{align}
As expected, PLog is the plethystic logarithm, that is, the inverse of plethystic exponential. For the theory we are discussing here, we find
\begin{align}
    \widetilde{I}_{(2,2,2)}=6t^4-3t^5\chi_{\bm{2}}(y)+3t^6-3t^7\chi_{\bm{2}}(y)-9t^8+7t^9\chi_{\bm{2}}(y)+O(t^{10})\,.
\end{align}
In fact, one can identify which short multiplets belonging to the 4d $\CN=1$ superconformal algebra contribute to each term of the superconformal index up to some low order in $t$. The contribution of each 4d $\CN=1$ short multiplet to the superconformal index is summarized in Appendix A of \cite{Gadde:2010en}, which we repeat with our notational conventions in Table \ref{tbl:multipletsci} of this paper. Up to $t^{r<8}$ order, we identify which short multiplets contribute to the reduced plethystic logarithm as follows: 
\begin{align}
    \widetilde{I}_{(2,2,2)}=6\overline{\mathcal{B}}_{\frac{4}{3}(0,0)}+3\overline{\mathcal{B}}_{\frac{5}{3}(\frac{1}{2},0)}+5\overline{\mathcal{B}}_{2(0,0)}+2\widehat{\mathcal{C}}_{(0,0)}+3\overline{\mathcal{B}}_{\frac{7}{3}(\frac{1}{2},0)}+O(t^8)\,.
\end{align}
However, the problem of determining the multiplet spectrum from the superconformal index does not have a unique solution; in this case, an ambiguity first arises at the order of $\mathcal{O}(t^8)$. There are four possible short multiplets
\begin{align}
    \overline{\mathcal{C}}_{-\frac{1}{3}(0,1)},\quad  \overline{\mathcal{C}}_{\frac{2}{3}(0,0)},\quad \mathcal{D}_{(0,1)},\quad -\overline{\mathcal{B}}_{\frac{8}{3}(0,0)}
\end{align}
that may contribute to the $-9t^8$ term from the index. Since $\mathcal{D}_{(0, 1)}$ comes from the higher-spin free field \cite{Cordova:2016emh}, it is absent for any interacting theory. 
Despite that we cannot give a full-proof rigorous argument, we give a heuristic reasoning why we think that there are no $\mathcal{C}$-type multiplets. This is because we find that from the refined index (see equation \eqref{eqn:refinedI222}) by turning on all the flavor fugacities, we see that the $t^8$ term naturally arises from the products of $\overline{\mathcal{B}}_{\frac{4}{3}(0, 0)}$, while the OPE of $\overline{\mathcal{B}}_{\frac{4}{3}(0, 0)}\times \overline{\mathcal{B}}_{\frac{4}{3}(0, 0)}$, the $\overline{\mathcal{B}}_{\frac{8}{3}(0,0)}$ multiplet appears but the $\overline{\mathcal{C}}_{-\frac{1}{3}(0,1)},\overline{\mathcal{C}}_{\frac{2}{3}(0,0)}$ multiplets are not present. As the six operators that contribute at order $t^4$ are of the form
\begin{equation}
  Q^2u_i \quad\text{and}\quad \operatorname{Tr}\mu_i\mu_{j\neq i}\quad (i,j\in\{1,2,3\}),
\end{equation}
we would naively expect that there exist $21$ operators at order $t^8$:
\begin{equation}
  (Q^2u_i)(Q^2u_j) , \qquad (Q^2u_i)\operatorname{Tr}\mu_j\mu_{k\neq j} , \qquad (\operatorname{Tr}\mu_i\mu_{j\neq i})^2 .
\end{equation}
However, we know that $(Q^2u_i)^2$ from each $\mathcal{D}_2(SU(3))$ theory are absent, and we expect that $(Q^2u_i)\operatorname{Tr}\mu_j\mu_{k\neq j}$ only exists if $i$, $j$, and $k$ are all distinct, since otherwise it would behave like a mixed Coulomb--Higgs operator of one of the $\mathcal{D}_2(SU(3))$ building blocks. Altogether, this gives $9$ relations at order $t^8$, which is reflected in the $-9t^8$ term in the plethystic log. Therefore, we claim that the $-9 t^8$ term appears in the index comes entirely from $-\overline{\mathcal{B}}_{\frac{8}{3}(0,0)}$. Hence, while naively there would be $6 \times 7/2 = 21$ $\overline{\mathcal{B}}_{\frac{8}{3}(0,0)}$ multiplets in the theory, there are only $12$ of them present.

\subsection{Gauging 4 copies of \texorpdfstring{$\mathcal{D}_2(SU(3))$}{D2(SU3)}}

In an analogous manner, we find that the reduced index of the theory constructed via gluing four copies of $\mathcal{D}_2(SU(3))$ by $\mathcal{N}=1$ gauging is
\begin{align}\label{eqn:i2222}
    \widehat{I}_{(2,2,2,2)}=t^{\frac{9}{2}}\left(8-4\chi_{\bm{2}}(y)\right)+3t^6+t^9\left(46-35\chi_{\bm{2}}(y)+6\chi_{\bm{3}}(y)\right)+O(t^{\frac{21}{2}}) \,.
\end{align}
There are six marginal operators of the form $\trace \mu_i\mu_j$, where we recall that the $\trace \mu_i^2$ operators are projected out by the chiral ring relation of $\mathcal{D}_2(SU(3))$, and there are three non-anomalous $U(1)$ flavor symmetries from $\mathcal{F}_i-\mathcal{F}_{i+1}$; thus, we expect that the contribution to the $t^6$ term is $6-3=3$, which agrees with equation \eqref{eqn:i2222}. The $t^7\chi_{\bm{2}}(y)v^2$ term in the index of each $\mathcal{D}_2(SU(3))$ contributes to the index of the gauged theory as the term $4t^6\chi_{\bm{2}}(y)$; similarly to the gauging of three copies of $\mathcal{D}_2(SU(3))$, this is canceled by the contribution from the four operators of the form $\operatorname{Tr}\lambda\mu_i$. There are eight relevant operators, among which four of them are the Coulomb branch operators of each individual $\mathcal{D}_2(SU(3))$ theory. The other four relevant operators are the $\mathcal{N}=2$ superpartners of the Coulomb branch operators of each $\mathcal{D}_2(SU(3))$ theory before gauging.

We write the reduced plethystic log of the index as
\begin{align}\label{eqn:billygoat}
    \widetilde{I}_{(2,2,2,2)}=t^{\frac{9}{2}}(8-4\chi_{\bm{2}}(y))+3t^6-3t^9\chi_{\bm{2}}(y)+O(t^{{\frac{21}{2}}})\,.
\end{align}
At low orders in $t$, the superconformal multiplets generating the chiral ring can be determined unambiguously from the index, and we find that it can be written as
\begin{align}
    \widetilde{I}_{(2,2,2,2)}=8\overline{\mathcal{B}}_{\frac{3}{2}(0,0)}+4\overline{\mathcal{B}}_{\frac{3}{2}(\frac{1}{2},0)}+6\overline{\mathcal{B}}_{2(0,0)}+3\widehat{\mathcal{C}}_{(0,0)}+O(t^9)\,.
\end{align}
In this expression, each of the listed short multiplets provides a shorthand for the plethystic logarithm of the contribution of that multiplet to the superconformal index, as summarized in Table \ref{tbl:multipletsci}. Thus it is easy to see that the spectrum of low-scaling dimension multiplets listed here reproduces the plethystic log in equation \eqref{eqn:billygoat}. 

\subsection{Gauging 5 copies of \texorpdfstring{$\mathcal{D}_2(SU(3))$}{D2(SU3)}}

At this point we expect that the reader is familiar with the procedure of using the expression in equation \eqref{eqn:plogreduced} to determine the reduced superconformal index. Then, without further ado, we find the reduced index of the theory comprised of five copies of $\mathcal{D}_2(SU(3))$ glued together by $\mathcal{N}=1$ gauging as
\begin{align}
    \begin{split}
    \widehat{I}_{(2,2,2,2,2)}=\ &5t^{\frac{18}{5}}-5t^{\frac{21}{5}}\chi_{\bm{2}}(y)+5t^{\frac{24}{5}}-4t^6+25t^{\frac{36}{5}}-25t^{\frac{39}{5}}\chi_{\bm{2}}(y)+t^{\frac{42}{5}}(40+10\chi_{\bm{3}}(y)) \\
    &-21t^9\chi_{\bm{2}}(y)-20t^{\frac{48}{5}}+O(t^{10}) \,.
    \end{split}
\end{align}
There are no marginal operators in this theory and four $U(1)$ flavor symmetries contributing $(-4)$ at the $t^6$ order. There are ten relevant scalar operators, where five of them are the Coulomb branch operators in the individual $\mathcal{D}_2(SU(3))$ theories and they contribute to the $5t^{\frac{18}{5}}$ term in the index. The other five relevant scalar operators correspond to the $5t^{\frac{24}{5}}$ term, and they arise from the $\mathcal{N}=2$ superdescendants of the Coulomb branch operators from each $\mathcal{D}_2(SU(3))$ theory. The contributions of all of the relevant and marginal operators are summarized in Table \ref{tbl:2s}.

To find the generators and relations of the chiral ring, we determine that the reduced plethystic log of the index is
\begin{align}
    \widetilde{I}_{(2,2,2,2,2)}=5t^{\frac{18}{5}}-5t^{\frac{21}{5}}+5t^{\frac{24}{5}}-4t^6+10t^{\frac{36}{5}}+4t^9-15t^{\frac{48}{5}}+O(t^{\frac{51}{5}})\,.
\end{align}
Up to $t^6$ order, we find that the following $\mathcal{N}=1$ supermultiplets contribute to the index:
\begin{align}
    \widetilde{I}_{(2,2,2,2,2)}=5\overline{\mathcal{B}}_{\frac{6}{5}(0,0)}+5\overline{\mathcal{B}}_{\frac{7}{5}(\frac{1}{2},0)}+5\overline{\mathcal{B}}_{\frac{8}{5}(0,0)}+4\widehat{\mathcal{C}}_{(0,0)}+O(t^{\frac{36}{5}}) \,.
\end{align}

\subsection{Gauging 6 copies of \texorpdfstring{$\mathcal{D}_2(SU(3))$}{D2(SU3)}}\label{sec:6D23}

Finally, we consider the SCFT obtained via the conformal $\mathcal{N}=1$ gauging of the $SU(3)$ flavor symmetry of six copies of the $\mathcal{D}_2(SU(3))$ theory. The reduced superconformal index of the resulting SCFT is
\begin{align}\label{eqn:ind222222}
    \begin{split}
        \widehat{I}_{(2,2,2,2,2,2)}=\ &6t^3-6t^4\chi_{\bm{2}}(y)+6t^5+16t^6-36t^7\chi_{\bm{2}}(y)+t^8\left(72+15\chi_{\bm{3}}(y)\right)\\
        &+t^9(26-16\chi_{\bm{2}}(y))+O(t^{10}) \,.
    \end{split}
\end{align}
There are five $U(1)$ flavor symmetries in this theory, each of which has a current contributing $(-1)$ to the $t^6$ term in the index. There are $21$ marginal operators obtained from the product of pairs of Coulomb branch operators: $u_iu_j$. All of the relevant operators contributing to this index are the six Coulomb branch operators and their $\mathcal{N}=2$ superpartners. For each of these gaugings of $\mathcal{D}_2(SU(3))$ theories, we have summarized the relevant and marginal operators that contribute to the index in Table \ref{tbl:2s}. 

Although it begins to become ambiguous, we can perform the operator spectroscopy also for the low R-charge irrelevant operators. We explore such an analysis in this example by explaining the operators that contribute to the $t^7$ and $t^8$ terms that appear in the index in equation \eqref{eqn:ind222222}. At $t^7$, there are $36$ operators of the form $u_i Q u_j$ that contribute the complete $-36\chi_{\bm{2}}(y)$. Putatively, at $t^7$, there also exist the operators $\trace \lambda \mu_i$, however, as discussed in Section \ref{sec:3D23}, these recombine with the operators contributing $\chi_{\bm{2}}(y)t^7$ in each $\mathcal{D}_2(SU(3))$ theory to form long multiplets in the gauged theory; thus they do not contribute to the superconformal index. At $t^8$ the contributing operators are $u_i Q^2 u_j$, $Qu_i Qu_j |_{\bm{1}}$, $\trace \mu_i \mu_{j \neq i}$, and $Qu_i Qu_{j\neq i} |_{\bm{3}}$, where we observe from the $\mathcal{D}_2(SU(3))$ superconformal index in equation \eqref{eqn:Dpindices} that the putative operators $QuQu|_{\bm{3}}$ do not exist. The contributions from all of these operators reproduce the $t^8$ term in the index in equation \eqref{eqn:ind222222}.

We further find that the reduced plethystic log of the index is
\begin{align}
    \widetilde{I}_{(2,2,2,2,2,2)}=6t^3-6t^4\chi_{\bm{2}}(y)+6t^5-5t^6-15t^8-5t^9\chi_{\bm{2}}(y)+O(t^{10}) \,,
\end{align}
and we can also determine the short $\mathcal{N}=1$ multiplets that contribute up to order $t^8$ as follows:
\begin{align}
   \widetilde{I}_{(2,2,2,2,2,2)}=6\overline{\mathcal{B}}_{1(0,0)}+6\overline{\mathcal{B}}_{\frac{4}{3}(\frac{1}{2},0)}+6\overline{\mathcal{B}}_{\frac{5}{3}(0,0)}+5\widehat{\mathcal{C}}_{(0,0)}+O(t^8)\,. 
\end{align}
It is straightforward to see that the superconformal primaries of the $\overline{\mathcal{B}}_{1(0,0)}$ multiplets are the $u_i$, of the $\overline{\mathcal{B}}_{\frac{4}{3}(\frac{1}{2},0)}$ multiplets are the $Qu_i$, and of the $\overline{\mathcal{B}}_{\frac{5}{3}(0,0)}$ multiplets are the $Q^2u_i$. The five $\widehat{\mathcal{C}}_{(0,0)}$ multiplets, which have scalar fields as their superconformal primaries, contain the five $U(1)$ flavor currents. This matches precisely with the known decompositions of the $\mathcal{N}=2$ $\mathcal{E}$-type multiplets into $\mathcal{N}=1$ superconformal multiplets, as given in equation \eqref{eqn:EtoB}. 

\section{\texorpdfstring{\boldmath{$\mathcal{N}=1$}}{N=1} gaugings of multiple \texorpdfstring{\boldmath{$\mathcal{D}_3(SU(2))$}}{D3(SU2)}}
\label{sec:multipleD3SU2}

Next, we turn to the study of a simple class of theories with $G=SU(2)$. We consider several theories built out of between two and four copies of the $\mathcal{D}_3(SU(2))$ theory via $\mathcal{N}=1$ gauging. As in the previous section, this construction will generally give rise to non-Lagrangian theories as $p$ and $N$, i.e., $3$ and $2$, are coprime. Each index demonstrates the absence of non-unitary contributions and the relevant and marginal operators that contribute to these three indices are summarized in Table \ref{tbl:3s}. 

\subsection{Gauging 2 copies of \texorpdfstring{$\mathcal{D}_3(SU(2))$}{D3(SU2)}}

Two copies of $\mathcal{D}_3(SU(2))$ can be glued together via $\mathcal{N}=1$ gauging to give rise to a theory with the following quiver diagram:
\begin{align*}
\begin{aligned}
    \begin{tikzpicture}
      \node[gaugeN1,ball color = gray!10, opacity = 0.8] (s0) {$SU(2)$};
      \node[d2] (c1) [left=0.6cm of s0] {$\mathcal{D}_{3}(SU(2))$};
      \node[d2] (c2) [right=0.6cm of s0] {$\mathcal{D}_{3}(SU(2))$};
      \draw (s0.east) -- (c2.west);
      \draw (s0.west) -- (c1.east);
    \end{tikzpicture} 
\end{aligned}\,.
\end{align*}
The reduced index of this theory with two copies of $\mathcal{D}_3(SU(2))$ can be computed again by an application of the integral formula given in equation \eqref{eqn:swipernoswiping}. We find that  
\begin{align}
    \begin{split}
        \widehat{I}_{(3,3)}=3t^3-2t^{\frac{9}{2}}\chi_{\bm{2}}(y)+3t^6-2t^{\frac{15}{2}}\chi_{\bm{2}}(y)+t^9\left(3+2\chi_{\bm{2}}(y)\right)+O\left(t^{\frac{21}{2}}\right) \,.
    \end{split}
\end{align}
The operators associated to each of these terms can be determined straightforwardly, as described in the previous section, and we do not belabor the point here; the operators are summarzied in Table \ref{tbl:3s}. After taking the plethystic log and removing the contributions from the conformal descendants, then we can see that it is written as
\begin{align}
    \widetilde{I}_{(3,3)}=3t^3-2t^{\frac{9}{2}}\chi_{\bm{2}}(y)-3t^6+4t^{\frac{15}{2}}\chi_{\bm{2}}(y)-t^9(1+\chi_{\bm{5}}(y))+O(t^{\frac{21}{2}})\,.
\end{align}
From this expression we can perform the $\mathcal{N}=1$ multiplet spectroscopy unambiguously up to order $t^6$, and we find that the reduced plethystic log can be written as
\begin{align}
    \widetilde{I}_{(3,3)}=3\overline{\mathcal{B}}_{1(0,0)}+2\overline{\mathcal{B}}_{\frac{3}{2}(\frac{1}{2},0)}-2\overline{\mathcal{B}}_{2(0,0)}+\widehat{\mathcal{C}}_{(0,0)}+O(t^{\frac{15}{2}}) \,,
\end{align}
where we note again that we have abused notation and wrote the name of the supermultiplet as a shorthand for the contribution to the reduced plethystic log of the superconformal index from that multiplet, as summarized in Table \ref{tbl:multipletsci}. 

The term $-2\overline{\mathcal{B}}_{2(0,0)}$ comes from two separate contributions: firstly, we get +2 from the two Coulomb branch operators $(u_1, u_2)$ in each $\mathcal{D}_3 (SU(2))$. The other (negative) contribution to this term comes from non-trivial relations between the chiral primaries in the operator product expansion of $\overline{\mathcal{B}}_{1(0,0)}\times\overline{\mathcal{B}}_{1(0,0)}$. More precisely, the superconformal primaries belonging to the  $6\overline{\mathcal{B}}_{2(0,0)}$ multiplets appearing in the OPE $3\overline{\mathcal{B}}_{1(0,0)}\times_{\text{sym}}3\overline{\mathcal{B}}_{1(0,0)}$ are, naively
\begin{align*}
    (Q^2u_1)(Q^2u_1), (Q^2u_2)(Q^2u_2), (Q^2u_1)\operatorname{Tr}\mu_1\mu_2, (Q^2u_2)\operatorname{Tr}\mu_1\mu_2, (Q^2u_1)(Q^2u_2) , (\operatorname{Tr}\mu_1\mu_2)^2 \,.
\end{align*} 
The first four are lifted by the chiral ring relations. This is deduced by looking at the refined index, given in equation \eqref{eq:I33ref}. Hence, these chiral ring relations contribute $-4\overline{\mathcal{B}}_{2(0,0)}$. Thus, we end up with $-2\overline{\mathcal{B}}_{2(0,0)}$.  

\subsection{Gauging 3 copies of \texorpdfstring{$\mathcal{D}_3(SU(2))$}{D3(SU2)}}

Now we consider the theories constructed from three copies of the $\mathcal{D}_{3}(SU(2))$ theory via $\mathcal{N}=1$ gauging, whose quiver diagram is given by 
\begin{align*}
\begin{aligned}
    \begin{tikzpicture}
      \node[gaugeN1,ball color = gray!10, opacity = 0.8] (s0) {$SU(2)$};
      \node[d2] (c1) [left=0.6cm of s0] {$\mathcal{D}_{3}(SU(2))$};
      \node[d2] (c2) [right=0.6cm of s0] {$\mathcal{D}_{3}(SU(2))$};
      \node[d2] (c3) [above=0.4cm of s0] {$\mathcal{D}_{3}(SU(2))$};
      \draw (s0.east) -- (c2.west);
      \draw (s0.west) -- (c1.east);
      \draw (s0.north) -- (c3.south);
    \end{tikzpicture} 
\end{aligned}\,.
\end{align*}
The reduced index of three copies of $\mathcal{D}_3(SU(2))$ glued together by $\mathcal{N}=1$ gauging is
\begin{align}
    \begin{split}
        \widehat{I}_{(3,3,3)}=\
        &t^4\left(6-3\chi_{\bm{2}}(y)\right)+t^6+t^8\left(21-15\chi_{\bm{2}}(y)+3\chi_{\bm{3}}(y)\right)+t^9\left(1-\chi_{\bm{2}}(y)\right)+O(t^{10}) \,.
    \end{split}
\end{align}
We can see that $\widehat{I}_{(3,3,3)}$ evinces a one-dimensional conformal manifold, and the relevant and marginal operators are listed in Table \ref{tbl:3s}. We also list the reduced plethystic log and express the short multiplets that contribute to the index at low orders:
\begin{align}
    \widetilde{I}_{(3,3,3)}=&\,t^4(t-3\chi_{\bm{2}}(y))+t^6+t^8(-6+3\chi_{\bm{2}}(y))+t^9(1-\chi_{\bm{2}}(y))+O(t^{10})\\
    =&\,6\overline{\mathcal{B}}_{\frac{4}{3}(0,0)}+3\overline{\mathcal{B}}_{\frac{4}{3}(\frac{1}{2},0)}+3\overline{\mathcal{B}}_{2(0,0)}+2\widehat{\mathcal{C}}_{(0,0)}+O(t^8)\,.
\end{align}

\subsection{Gauging 4 copies of \texorpdfstring{$\mathcal{D}_3(SU(2))$}{D3(SU2)}}

We further consider the theories constructed with four copies of the $\mathcal{D}_{3}(SU(2))$ theory glued via $\mathcal{N}=1$ gauging. The corresponding quiver diagram is
\begin{align*}
\begin{aligned}
    \begin{tikzpicture}
      \node[gaugeN1,ball color = gray!10, opacity = 0.8] (s0) {$SU(2)$};
      \node[d2] (c1) [left=0.6cm of s0] {$\mathcal{D}_{3}(SU(2))$};
      \node[d2] (c2) [right=0.6cm of s0] {$\mathcal{D}_{3}(SU(2))$};
      \node[d2] (c3) [above=0.4cm of s0] {$\mathcal{D}_{3}(SU(2))$};
      \node[d2] (c4) [below=0.4cm of s0] {$\mathcal{D}_{3}(SU(2))$};
      \draw (s0.east) -- (c2.west);
      \draw (s0.west) -- (c1.east);
      \draw (s0.north) -- (c3.south);
      \draw (s0.south) -- (c4.north);
    \end{tikzpicture} 
\end{aligned}\,,
\end{align*}
and the reduced index of the theory thus obtained is 
\begin{align}
    \begin{split}
        \widehat{I}_{(3,3,3,3)}=\
        &4t^3-4t^\frac{15}{4}\chi_{\bm{2}}(y)+4t^\frac{9}{2}+7t^6-16t^\frac{27}{4}\chi_{\bm{2}}(y)\\
        &+t^\frac{15}{2}\left(28+6\chi_{\bm{3}}(y)\right)+t^\frac{33}{4}\left(4-12\chi_{\bm{2}}(y)\right)+t^9\left(14+5\chi_{\bm{2}}(y)\right)\\
        &-t^\frac{39}{4}\left(20+28\chi_{\bm{2}}(y)+16\chi_{\bm{3}}(y)\right)+O(t^{10}) \,.
    \end{split}
\end{align}
We compute the reduced plethystic log of the index $\widetilde{I}_{(3,3,3,3)}$ in order to exhibit the generators and relations of the chiral ring:
\begin{align}
\begin{split}
    \widetilde{I}_{(3,3,3,3)}=&\,4t^3-4t^{\frac{15}{4}}\chi_{\bm{2}}(y)+4t^{\frac{9}{2}}-3t^6+2t^{\frac{15}{2}}+t^{\frac{33}{4}}(4+4\chi_{\bm{2}}(y))\\
    &-t^9(4+\chi_{\bm{2}}(y))-4t^{\frac{33}{4}}+O(t^{\frac{41}{4}})\,.
    \end{split}
\end{align}
We can also identify every short multiplet containing relevant and marginal operators that contributes to the superconformal index as follows:
\begin{align}
    \widetilde{I}_{(3,3,3,3)}=4\overline{\mathcal{B}}_{1(0,0)}+4\overline{\mathcal{B}}_{\frac{5}{4}(\frac{1}{2},0)}+4\overline{\mathcal{B}}_{\frac{3}{2}(0,0)}+3\widehat{\mathcal{C}}_{(0,0)}+O(t^{\frac{15}{2}})\,.
\end{align}
As expected, this is consistent with the decomposition of the $\mathcal{N}=2$ $\mathcal{E}$-type multiplets containing the low-scaling dimension Coulomb branch operators of each of the $\mathcal{D}_3(SU(2))$ building blocks, as in equation \eqref{eqn:EtoB}, combined with the non-anomalous $U(1)$ flavor currents that survive from the $\mathcal{N}=2$ R-symmetry after $\mathcal{N}=1$ gauging.

\section{\texorpdfstring{\boldmath{$\mathcal{N}=1$}}{N=1} SCFT constructions with \texorpdfstring{\boldmath{$\mathcal{D}_5(SU(2))$}}{D5(SU2)}}
\label{sec:indexirrational}

In this section, we consider examples where the gauging involves at least one copy of the $\mathcal{D}_5(SU(2))$ theory. This is the theory with the largest value of $p$ for which the superconformal index can be computed in a reasonable timeframe. The first example, in Section \ref{sec:335}, is the first instance where we determine the index for gaugings involving differing $p_i$, and as such we find that the superconformal R-symmetry involves a mixing with irrational coefficients. In the second example, we consider gauging together two copies of $\mathcal{D}_5(SU(2))$. The resulting operators spectroscopy in these two cases is summarized in Table \ref{tbl:5s}.

\subsection{Gauging 2 copies of \texorpdfstring{$\mathcal{D}_3(SU(2))$}{D3(SU2)} and one \texorpdfstring{$\mathcal{D}_5(SU(2))$}{D5(SU2)}}\label{sec:335}

We consider the theory composed via gluing two copies of $\mathcal{D}_3(SU(2))$ and a single $\mathcal{D}_5(SU(2))$ together, via $(\mathcal{N}=1)$-gauging, which can be depicted as
\begin{align}
\begin{aligned}
    \begin{tikzpicture}
      \node[gaugeN1,ball color = gray!10, opacity = 0.8] (s0) {$SU(2)$};
      \node[d2] (c1) [left=0.6cm of s0] {$\mathcal{D}_{3}(SU(2))$};
      \node[d2] (c2) [right=0.6cm of s0] {$\mathcal{D}_{3}(SU(2))$};
      \node[d2] (c3) [above=0.4cm of s0] {$\mathcal{D}_{5}(SU(2))$};
      \draw (s0.east) -- (c2.west);
      \draw (s0.west) -- (c1.east);
      \draw (s0.north) -- (c3.south);
    \end{tikzpicture} \,.
\end{aligned}
\end{align}
We find the reduced superconformal index of this theory is
\begin{align}
    \begin{split}
        \widehat{I}_{(3,3,5)}=\
        &t^{3.05814}-t^{3.50969}\chi_{\bm{2}}(y)+3t^{3.96124}-2t^{3.99031}\chi_{\bm{2}}(y)+2t^{4.01938}\\
        &+t^{4.07752}-t^{4.52907}\chi_{\bm{2}}(y)+t^{4.98062}-2t^6+t^{6.05814}+t^{6.11627}+O(t^{6.48062}) \,.
    \end{split}
\end{align}
This theory is an example where the mixing coefficients $\epsilon_i$ are irrational, and thus the theory has \emph{irrational} charges, as seen from the irrational powers of $t$; however, we can see that all the terms that appear are consistent with unitarity. For this theory, we study some of the operators that contribute to the superconformal index. Recall that each copy of the $\mathcal{D}_3(SU(2))$ theory has a single Coulomb branch operator, which we call $u_1$ and $u_2$, and the $\mathcal{D}_5(SU(2))$ theory has two Coulomb branch operators, $u_3$ and $v_3$. We find that $u_3$ contributes to the $t^{3.05814}$ term; $u_1$ and $u_2$ contribute to the $t^{3.96124}$ term; and $v_3$ provides the $t^{4.07752}$ term. All other relevant terms are provided by the $\mathcal{N}=2$ superdescendants of these four Coulomb branch operators. There are two $U(1)$ flavor currents, and no marginal operators, that contribute to the $t^6$ term. Finally, we can see that $\trace \mu_1 \mu_2$, which is an irrelevant operator, contributes to the $t^{6.05814}$ term. 

Up to $O(t^{6.05814})$ the reduced index is identical with reduced plethystic log of the index since the first composite operator appears at $O(t^{6.11627})$. Every relevant operator captured by the superconformal index is actually a superconformal primary operator of a $\overline{\mathcal{B}}$-type $\mathcal{N}=1$ superconformal  multiplet. Here, we exhibit the short multiplets that are captured by the index up to order $t^6$:
\begin{align}
   \begin{split} \widetilde{I}_{(3,3,5)}=&\,\overline{\mathcal{B}}_{1.01938(0,0)}+\overline{\mathcal{B}}_{1.16990(\frac{1}{2},0)}+3\overline{\mathcal{B}}_{1.32041(0,0)}+2\overline{\mathcal{B}}_{1.33010(\frac{1}{2},0)}+2\overline{\mathcal{B}}_{1.33979(0,0)}\\
   &+\overline{\mathcal{B}}_{1.35917(0,0)}+\overline{\mathcal{B}}_{1.50969(\frac{1}{2},0)}+\overline{\mathcal{B}}_{1.66021(0,0)}+2\widehat{C}_{(0,0)}+O(t^{6.05814})\,.
   \end{split}
\end{align}
These supermultiplets and their associated primary operators are listed in Table \ref{tbl:5s}.

\subsection{Gauging 2 copies of \texorpdfstring{$\mathcal{D}_5(SU(2))$}{D5(SU2)}}

Another theory we consider which involves the $\mathcal{D}_{5}(SU(2))$ building block is the theory obtained via gauging two copies of the $\mathcal{D}_{5}(SU(2))$ theory. The resulting SCFT can be written as the quiver
\begin{align}
\begin{aligned}
    \begin{tikzpicture}
      \node[gaugeN1,ball color = gray!10, opacity = 0.8] (s0) {$SU(2)$};
      \node[d2] (c1) [left=0.6cm of s0] {$\mathcal{D}_{5}(SU(2))$};
      \node[d2] (c2) [right=0.6cm of s0] {$\mathcal{D}_{5}(SU(2))$};
      \draw (s0.east) -- (c2.west);
      \draw (s0.west) -- (c1.east);
    \end{tikzpicture}
\end{aligned} \,.
\end{align}
By computing the superconformal index we can verify that there are no terms that violate unitarity and thus confirm that we obtain an interacting SCFT with $a = c$ in the infrared. The reduced index of this theory is 
\begin{align}
    \begin{split}
        \widehat{I}_{(5,5)}=\
        &2t^3-2t^{\frac{15}{4}}\chi_{\bm{2}}(y)+5t^{\frac{9}{2}}-2t^{\frac{21}{4}}\chi_{\bm{2}}(y)+2t^6-2t^{\frac{27}{4}}\chi_{\bm{2}}(y)+O(t^7) \,,
    \end{split}
\end{align}
which has rational exponents, as expected since the mixing coefficients are themselves rational. We call the two Coulomb branch operators of $\mathcal{D}_5(SU(2))$ as $u$ and $v$, and they have dimensions $6/5$ and $8/5$, respectively. In the gauged theory, we know that there exist two marginal operators that come from $v_1$ and $v_2$. There also exists a marginal operator $Q^2u_1Q^2u_2$, however we can see from the superconformal index of the $\mathcal{D}_5(SU(2))$ theory, given in equation \eqref{eqn:Dpindices}, that the putative $(Q^2u_i)^2$ operators do not contribute to the index. There is a single non-anomalous $U(1)$ flavor symmetry. The current for this flavor symmetry and the three marginal operators contribute to the coefficient $3-1=2$ of the $t^6$ term. The relevant terms in the index are contributed to by the following operators: $t^3$ is $Q^2 u_i$, $t^{15/4}$ is $Qu_i$, $t^{9/2}$ is $u_i$, $Q^2 v_i$, and $\trace \mu_1 \mu_2$, and finally $t^{21/4}$ is $Qv_i$. We summarize this operator content, together with the relevant and marginal operators in terms of $\mathcal{N}=1$ superconformal multiplets, as determined from the reduced plethystic logarithm, in Table \ref{tbl:5s}.

\section{\texorpdfstring{\boldmath{$\mathcal{N}=1$}}{N=1} theories with adjoint chirals}
\label{sec:indexadjoints}

We have now determined the superconformal indices for a variety of the infrared SCFTs with $a = c$ that we constructed in \cite{Kang:2021ccs}. We now turn to examples where, in addition to the gauged $\mathcal{D}_p(G)$ theories, we also include one or two extra adjoint chiral multiplets. The superconformal index of this SCFT can be determined using the expression for the superconformal index of the $\mathcal{D}_p(G)$ theory, as in equation \eqref{eqn:Dpindices}, and the known expression for the index of a weakly-coupled chiral multiplet. A chiral multiplet in the representation $(\textbf{adj},\bm{R})$ of a flavor symmetry $G\times \widetilde{F}$ has index
\begin{equation}\label{eqn:chiralindcontrib}
    I_\text{adjoint chiral} = \operatorname{PE}\left[\frac{t^{3R_\phi}\chi_{\widetilde{F},\bm{R}}(\bm{v})-t^{6-3R_\phi}\chi_{\widetilde{F},\overline{\bm{R}}}(v)}{(1-t^3y)(1-t^3/y)}\chi_{G,\textbf{adj}}(\bm{z})\right] \,,
\end{equation}
where $R_\phi$ is the $R$-charge of adjoint chiral $\phi$, $\bm{z}$ denotes the $G$ flavor fugacity that will be gauged together with $\mathcal{D}_p(G)$, and $v$ stands for the fugacities of the other flavor symmetry $\widetilde{F}$ collectively.

\subsection{Gauging 2 copies of \texorpdfstring{$\mathcal{D}_2(SU(3))$}{D2(SU3)} with an adjoint chiral}

As a first example, consider the SCFT constructed via the $(\mathcal{N}=1)$-gauging of two copies of the $\mathcal{D}_2(SU(3))$ theory, together with an additional adjoint chiral multiplet $\phi$:
\begin{align}
    \begin{aligned}
    \begin{tikzpicture}
      \node[gaugeN1,ball color = gray!10, opacity = 0.8] (s0) {$SU(3)$};
      \node[d2] (c2) [left=0.6cm of s0] {$\mathcal{D}_{2}(SU(3))$};
      \node[d2] (c3) [right=0.6cm of s0] {$\mathcal{D}_{2}(SU(3))$};
      \draw (s0.west) -- (c2.east);
      \draw (s0.east) -- (c3.west);
       \draw[dashed, ->] (s0) to[out=130, in=410, looseness=4] (s0);
    \end{tikzpicture} \,.
\end{aligned}
\label{eqn:1adj22}
\end{align}
The reduced index of this theory is straightforwardly worked out from the superconformal indices of the building blocks, and we find
\begin{align}
\begin{split}
    \widehat{I}_{(2,2),\bm{8}}^{n_a=1}=\
    &t^{2.5359}+3t^{3.80385}-3t^{4.26795}\chi_{\bm{2}}(y)+4t^{4.73205}+t^{5.0718}\\
    &-t^{5.5359}\chi_{\bm{2}}(y)+3t^{6.33975}+O(t^{6.80385}) \,.
    \end{split}
\end{align}
Here we utilized the notation $n_a$ to denote the number of adjoint chiral multiplets. We can see that the index for this theory contains no unitarity-violating terms, demonstrating that it does indeed flow to an infrared SCFT with $a=c$. The relevant and marginal operators that each of the terms in the index arise from are listed in Table \ref{tbl:2s}; similarly, it is straightforward to use the reduced plethystic logarithm to determine the $\mathcal{N}=1$ superconformal multiplets that contribute to the index at low orders, and these are also contained in Table \ref{tbl:2s}. 

\subsection{Gauging 1 copy of \texorpdfstring{$\mathcal{D}_3(SU(2))$}{D3(SU2)} with two adjoint chirals}

Next, we consider an example of a gauging involving two additional adjoint chiral multiplets, corresponding to the quiver
\vspace{-10mm}
\begin{align}
\begin{aligned}
    \begin{tikzpicture}
      \node[gaugeN1,ball color = gray!10, opacity = 0.8] (s0) {$SU(2)$};
      \node[d2] (c2) [left=0.6cm of s0] {$\mathcal{D}_{3}(SU(2))$};
      \draw (s0.west) -- (c2.east);
       \draw[dashed, ->] (s0) to[out=40, in=320, looseness=3] (s0);
      \draw[dashed, ->] (s0) to[out=56, in=304, looseness=7] (s0);
    \end{tikzpicture}
\end{aligned}\,.
\end{align}
\vspace{-16mm}

\noindent We find that the reduced index of this theory, constructed with a $\mathcal{D}_{3}(SU(2))$ gauged together with two adjoint chiral multiplets $\phi_1$ and $\phi_2$ attached, is
\begin{align}\label{eqn:SCI2chiral}
\begin{split}
    \widehat{I}_{(3),\bm{3},\bm{3}}^{n_a=2} =&t^{2.74273}+t^{3.68568}(3-\chi_{\bm{2}}(y))+t^{4.62864}-2t^{4.84284}\chi_{\bm{2}}(y)+t^{5.48545}\\
    &+2t^{5.7858}-4t^6+t^{6.42841}(3-\chi_{\bm{2}}(y))+O(t^{7.62864}) \,. 
    \end{split}
\end{align}
We can see that this index contains no unitarity violating terms, and thus we have a good infrared SCFT with $a = c$, as expected from the analysis of a subset of the protected operators in \cite{Kang:2021ccs}. The flavor symmetry of this theory is $U(1) \times SU(2)$, which can be seen in the $t^6$ term
\begin{equation}
-4t^6\ \rightarrow\ -(1+\chi_{\mathfrak{su}_2,\bm{3}}(v))t^6 \,,
\end{equation}
if we revive the $SU(2)$ flavor fugacity, $v$. The flavor symmetry is the anomaly-free part of the classical $U(1)\times U(1)\times SU(2)$ flavor symmetry, where the first $U(1)$ is the flavor remnant of $\mathcal{N}=2$ R-symmetry, and the remaining $U(1)\times SU(2)$ is the symmetry rotating the two adjoint chirals $\phi_{1,2}$. The relevant and marginal operators that contribute to the superconformal index are listed in Table \ref{tbl:3s}, and we now briefly describe their identification. Under the superconformal R-symmetry the R-charges of the adjoint chirals, the moment map, and the single Coulomb branch operator of the Argyres--Douglas theory are
\begin{equation}
    \begin{aligned}
        R(\phi_1) = R(\phi_2) &= \frac{87 - \sqrt{354}}{111} \sim 0.61428 \,, \\
        R(\mu) &= \frac{11 + 2\sqrt{354}}{37} \sim 1.31432 \,, \\
        R(u) &= \frac{252 - 8\sqrt{354}}{111} \sim 0.914242 \,.
    \end{aligned}
\end{equation}
If we consider operators built out of these objects then we find that the following are relevant scalar operators
\begin{equation}\label{eqn:rele}
    \begin{gathered}
        \trace \phi_1^2 \,, \quad \trace \phi_1 \phi_2 \,, \quad \trace \phi_2^2 \,, \quad\trace \mu \phi_1 \,, \quad \trace \mu \phi_2 \,, \quad u \,, \quad u^2 \,.
    \end{gathered}
\end{equation}
Each operator contributes a term $t^{3R}$, where $R$ is the R-charge of the operator, to the superconformal index, and thus we can see the following contributions: the Coulomb branch operator $u$ to $t^{2.74273}$, the three $\trace \phi_i \phi_j$ to $t^{3.68568}$, the two $\trace \mu \phi_i$ to $t^{5.7858}$, and the $u^2$ to $t^{5.48545}$. The theory has no marginal operators. The four operators $\trace \phi_i \phi_j \phi_k$ would a priori appear to be relevant operators, however they are absent in this specific case due to the absence of a cubic Casimir for $G = SU(2)$. In \cite{LANDSCAPE}, we consider $G = SU(N)$ and study the SCFTs obtained by renormalization group flows triggered by both the relevant operators in equation \eqref{eqn:rele} and the cubic operators, when they exist. These operators organize themselves into $\mathcal{N}=1$ superconformal multiplets, and the specific multiplets that contain the relevant and marginal operators can easily be determined from the plethystic log; we summarize the operators and their associated superconformal multiplets in Table \ref{tbl:3s}. 

\subsection{Gauging 1 copy of \texorpdfstring{$\mathcal{D}_5(SU(2))$}{D5(SU2)} with two adjoint chirals}
\label{sec:1D5}

We consider another example of a gauging with two adjoint chiral multiplets where the $\mathcal{D}_3(SU(2))$ from the previous section is replaced with the $\mathcal{D}_5(SU(2))$ Argyres--Douglas theory. The ultraviolet depiction of the resulting $\mathcal{N}=1$ SCFT is given by the following quiver
\vspace{-8mm}
\begin{align}
\begin{aligned}
    \begin{tikzpicture}
      \node[gaugeN1,ball color = gray!10, opacity = 0.8] (s0) {$SU(2)$};
      \node[d2] (c2) [left=0.6cm of s0] {$\mathcal{D}_{5}(SU(2))$};
      \draw (s0.west) -- (c2.east);
       \draw[dashed, ->] (s0) to[out=40, in=320, looseness=3] (s0);
      \draw[dashed, ->] (s0) to[out=56, in=304, looseness=7] (s0);
    \end{tikzpicture}
\end{aligned}\,.
\end{align}
\vspace{-13mm}

\noindent Using the by-now-familiar techniques, we find that the reduced superconformal index of the infrared SCFT is
\begin{align}
\begin{split}
\widehat{I}^{n_a=2}_{(5),\bm{3},\bm{3}}=\
&t^{2.42423}+t^{3.23231}-t^{3.40404}\chi_{\bm{2}}(y)+3t^{3.80808}-t^{4.21212}\chi_{\bm{2}}(y)+t^{4.38384}\\
&+t^{4.84847}-2t^{4.90404}\chi_{\bm{2}}(y)+t^{5.19192}+t^{5.65654}-t^{5.82827}\chi_{\bm{2}}(y)+2t^{5.88384}\\
&-4t^6+O\left(t^{6.23231}\right) \,.
\end{split}
\end{align}
The theory has the same $U(1)\times SU(2)$ flavor symmetry as the theory studied in the previous subsection. We can figure out the relevant and marginal operators that contribute to the index with the data of the infrared R-charges of each fields. The R-charges of the adjoint chiral multiplets, the moment map, and the Coulomb branch operator of the $\mathcal{D}_5(SU(2))$ theory are
\begin{align}
    \begin{split}
        R(\phi_1)=R(\phi_2)=\ &\frac{82-\sqrt{298}}{102}\sim 0.634680\,,\\
        R(\mu)=\ &\frac{49 + 5 \sqrt{298}}{102}\sim 1.32660\,,\\
        R(u)=\ &\frac{31-\sqrt{298}}{17}\sim 0.808078\,.
    \end{split}
\end{align}
We find that there are no terms corresponding to operators that violate the unitarity bound, thus we expect the IR theory in this example indeed has identical central charges. The full spectrum of relevant and marginal operators we find from index, and the $\mathcal{N}=1$ multiplets to which they belong, are listed in Table \ref{tbl:5s}. 

\subsection{Gauging 0 copies of \texorpdfstring{$\mathcal{D}_p(G)$}{Dp(G)} with two adjoint chirals}\label{sec:noDp}

Throughout this paper, and in our previous work \cite{Kang:2021ccs}, we focus on 4d $\mathcal{N}=1$ SCFTs that are built out of a diagonal gauging of Argyres--Douglas $\mathcal{D}_p(G)$ theories, together with the possible inclusion of adjoint-valued chiral multiplets. In fact, such a construction can lead to $\mathcal{N}=1$ theories with $a=c$ even if the number of 
Argyres--Douglas theories included as building blocks is zero! In this section, we consider the Lagrangian theory obtained 
formally by gauging the $G$ flavor symmetries of zero copies of any $\mathcal{D}_p(G)$ theory, together with two adjoint chiral multiplets. This is thus simply a quiver gauge theory with a gauge node $G$ and two adjoint chiral multiplets:
\vspace{-8mm}
\begin{align}
\begin{aligned}
    \begin{tikzpicture}
      \node[gaugeN1,ball color = gray!10, opacity = 0.8] (s0) {$G$};
       \draw[dashed, ->] (s0) to[out=40, in=320, looseness=3] (s0);
      \draw[dashed, ->] (s0) to[out=56, in=304, looseness=7] (s0);
    \end{tikzpicture}
\end{aligned}\,.
\end{align}
\vspace{-13mm}

\noindent The reader can easily confirm that this theory has $a=c$, for any value of $G$. It has an $SU(2)$ flavor symmetry whose fundamental representation rotates the two adjoint chiral multiplets, $\phi_1$ and $\phi_2$, while the classical $U(1)$ that rotates their phase is anomalous. We consider $G=SU(3)$, and then the reduced index of this theory can be determined utilizing the formula in equation \eqref{eqn:chiralindcontrib}. The result is simply written as 
\begin{align}
\begin{split}
    \widehat{I}^{n_a=2}_{\textbf{adj},\textbf{adj}}=\ &t^3\chi_{\mathfrak{su}_2,\bm{3}}(v)+t^{\frac{9}{2}}\left(\chi_{\mathfrak{su}_2,\bm{4}}(v)-\chi_{\mathfrak{su}_2,\bm{2}}(v)\chi_{\bm{2}}(y)\right)\\
    &+t^6\left(\chi_{\mathfrak{su}_2,\bm{5}}(v)-\chi_{\mathfrak{su}_2,\bm{3}}(v)+2-\chi_{\mathfrak{su}_2,\bm{3}}(v)\chi_{\bm{2}}(y)\right)+O\left(t^{\frac{15}{2}}\right)\,,
\end{split}
\end{align}
where we have refined the index by the fugacity $v$ of the $SU(2)$ flavor symmetry.
The theory contains seven relevant scalar operators 
\begin{align}\label{eqn:sabrina}
    \trace\phi_1^2\,,\quad\trace\phi_1\phi_2\,,\quad\trace\phi_2^2\,,\quad \trace\phi_1^3\,,\quad\trace\phi_1^2\phi_2\,,\quad\trace\phi_1\phi_2^2\,,\quad\trace\phi_2^3 \,,
\end{align} 
where the first three transform in the $\bm{3}$ of the $SU(2)$ flavor, and the latter four transform in the $\bm{4}$.
There are seven marginal scalar operators
\begin{align}\label{eqn:hilda}
\begin{split}
    \left(\trace\phi_1^2\right)^2, \trace\phi_1^2\,\trace\phi_1\phi_2, \trace\phi_1^2\,\trace\phi_2^2, \left(\trace\phi_1\phi_2\right)^2, \trace\phi_1\phi_2\,\trace\phi_2^2, \left(\trace\phi_2^2\right)^2,
    \trace [\phi_1 , \phi_2]^2 \,.
\end{split}
\end{align}
Among these seven marginal operators, four of them are exactly marginal and span a four-dimensional conformal manifold.

It turns out that $G=SU(3)$ is particularly special due to the absence of an independent quartic Casimir for that Lie algebra. When we have a larger gauge symmetry, we have more marginal operators. For example when $G=SU(4)$, the reduced index is
\begin{align}
\begin{split}
    \widehat{I}^{n_a=2}_{\textbf{adj},\textbf{adj}}=&\,t^3\chi_{\mathfrak{su2},\bf{3}}(v)+t^{\frac{9}{2}}\left(\chi_{\mathfrak{su}_2,\bm{4}}(v)-\chi_{\mathfrak{su}_2,\bm{2}}(v)\chi_{\bm{2}}(y)\right)\\
    &+t^6\left(2\chi_{\mathfrak{su}_2,\bm{5}}(v)-\chi_{\mathfrak{su}_2,\bm{3}}(v)+2-\chi_{\mathfrak{su}_2,\bm{3}}(v)\chi_{\bm{2}}(y)\right)+O\left(t^{\frac{15}{2}}\right)\,.
\end{split}
\end{align}
As we can see, there is an additional set of scalar marginal operators transforming in the $\bm{5}$ of the $SU(2)$ flavor. These are the operators
\begin{equation}\label{eqn:zelda}
    \trace \phi_1^4 \,,\quad \trace \phi_1^3 \phi_2 \,,\quad \trace \{\phi_1, \phi_2\}^2 \,,\quad \trace \phi_1 \phi_2^3 \,,\quad  \trace \phi_2^4 \,.
\end{equation}
These operators exist for the $a=c$ infrared SCFTs that have ultraviolet description as $SU(N)$ with two adjoint-valued chiral multiplets, for any $N > 3$. The landscape formed by superpotential deformations to new $\mathcal{N}=1$ SCFTs triggered by the operators in equations \eqref{eqn:sabrina}, \eqref{eqn:hilda}, and \eqref{eqn:zelda} has been studied in \cite{Intriligator:2003mi}. 

\section{Summary and future directions}
\label{sec:summary}

We have determined the superconformal indices for a wide variety of theories, and we have also discussed in some cases how we can analyze the computed indices to determine the low-dimension operator content of the theory; hence we are effectively performing \emph{operator spectroscopy}. We have verified that there is no unitary violating operator up to certain order, further supporting that our $a=c$ theories constructed are unitary interacting SCFTs. 
We now summarize the operators that contribute to the relevant and marginal terms up to $t^{\leq 6}$ for the indices of all of the $\mathcal{D}_p(G)$ gaugings discussed throughout this paper. (See Tables \ref{tbl:2s}, \ref{tbl:3s}, and \ref{tbl:5s}.)

The relevant and marginal operators typically fall into a few fixed categories. The first kind of operators that appear are those that arise from the $\mathcal{N}=2$ $\mathcal{E}$-type supermultiplets that contain the Coulomb branch operators. Let $Q$ denote the $\mathcal{N}=2$ supercharge which has non-zero $j_2$. The states $Qu$ and $Q^2u$ are super-descendants of the Coulomb branch operator $u$. After $\CN=1$ gauging, $Q$ is no longer a supercharge of the theory, and the states $u$, $Qu$, and $Q^2u$ are no longer related by supersymmetry; the are independent operators. Nevertheless, these states, and products of these states, regularly contribute relevant and marginal operators to the spectrum of these $a=c$ SCFTs. 

The second category of operators are those constructed out of the fields of the weakly-coupled content of the gauged theory. We variously include $\mathcal{N}=2$ vector multiplets, $\mathcal{N}=1$ vector multiplets, and $\mathcal{N}=1$ chiral multiplets. The fields with which we are concerned are the scalar field $\phi$ inside of the $\mathcal{N}=1$ chiral multiplet and the gaugino $\lambda$ inside of the $\mathcal{N}=1$ vector multiplet. When the gauging involves an $\mathcal{N}=2$ vector multiplet, as in the $\widehat{\Gamma}(G)$ theories, one introduces both an $\mathcal{N}=1$ vector and chiral multiplet, and the supercharge $Q$, charged under $j_2$, relates the gaugino and the scalar as $\lambda = Q \phi$. 

Finally, there are operators constructed out of the moment-maps of each $\mathcal{D}_p(G)$, $\mu$. Typically, the Joseph ideal in equation \eqref{eqn:joseph_ideal} removes operators containing $\mu^k$, and thus all contributions from the moment-maps involve products of moment maps from different $\mathcal{D}_p(G)$ origins. All non-cancelling contributions to the indices turn out to be formed either out of products of these three categories of operators, or else out of operators belonging to the flavor current multiplets.

In Table \ref{tbl:2s}, we list the relevant and marginal operator content, and how they contribute to the index, of the $\widehat{D}_4(SU(3))$ theory, the four theories obtained by $(\mathcal{N}=1)$-gauging of between three and six copies of $\mathcal{D}_2(SU(3))$, and the infrared theory obtained by $(\mathcal{N}=1)$-gauging of two copies of $\mathcal{D}_2(SU(3))$ together with a single additional adjoint chiral multiplet. Similarly, in Table \ref{tbl:3s}, we write the operator content for the $\widehat{E}_6(SU(2))$ theory, the $(\mathcal{N}=1)$-gaugings of three and four $\mathcal{D}_3(SU(2))$ theories, and finally the theory obtained via the $(\mathcal{N}=1)$-gauging of the $SU(2)$ flavor symmetry of a single $\mathcal{D}_3(SU(2))$ together with two additional adjoint-valued chiral multiplets. Lastly, in Table \ref{tbl:5s}, we write the relevant and marginal operators, the $\mathcal{N}=1$ superconformal multiplets that they belong to, and how they contribute to the index for each of the gauged theories that we consider involving a $\mathcal{D}_5(SU(2))$ building block.

We want to emphasize that the operator spectroscopy done in this paper has set the foundations to further study an even broader landscape of 4d SCFTs with $a=c$. In particular, we are exploring the landscape of 4d $\mathcal{N}=1$ SCFTs with $a=c$ by investigating if any superpotential deformation maintaining the $a=c$ property exists. In view of the analysis done in \cite{Intriligator:2003mi} on the superpotential deformations for SQCD with fundamental and adjoint chiral multiplets, we perform a similar analysis, and we find that the resulting 4d $\mathcal{N}=1$ SCFTs often preserve the $a=c$ property \cite{LANDSCAPE}.

Another important expectation is that analyzing the operator spectrum, as we have done in this paper, should be helpful for constructing potential supergravity dual theories to these 4d SCFTs with $a=c$. The fact that $a=c$ holds at finite $N$, where $N$ is the rank of the gauge algebra, requires a remarkable cancellation in the contributions to the subleading orders in any putative AdS$_5$ dual. Determining the precise mechanism which sources this cancellation is the subject of ongoing work. 
In the current paper, we have computed the superconformal indices only for low-rank theories; this is because the full index of the higher-rank $\mathcal{D}_p (G)$ theories are yet unavailable. It would be interesting to find a method to compute the index for arbitrary rank and look for the large $N$ behavior of the index, which should be helpful for understanding the holographic dual of these theories.\footnote{For the case of $\CN=2$ $\hat{\Gamma}(G)$ theories, the Schur index is available for arbitrary $N$ \cite{Bourdier:2015sga, Pan:2021mrw, Huang:2022bry, Hatsuda:2022xdv}.} The operator spectrum of these SCFTs, which we analyzed via their superconformal indices, provides constraints on the form of the supergravity duals, as these protected operators, and the renormalization group flows that they trigger, must be replicated in their holographic dual theories.

\begin{table}[H]
    \centering
    \footnotesize
    \renewcommand{\arraystretch}{1.1}
    \begin{threeparttable}
    \begin{tabular}{cccccc}
    \toprule
        \multirow{2}{*}{Index} & \multirow{2}{*}{Term} & \multirow{2}{*}{Positive} & \multirow{2}{*}{Negative} & \multirow{2}{*}{Total} & $\mathcal{N}=1$ \\
         & & & & & Multiplets \\\midrule
        \multirow{7}{*}{$\widehat{I}_{\widehat{D}_4(SU(3))}$} & $t^{3}v^{-3}$ & $u_i$ & -- & $4$ & $4\overline{\mathcal{B}}_{1(0,0)}$ \\
         & $t^{4}v^{-1}\chi_{\bm{2}}(y)$ & -- & $Qu_i$ & $-4$ & $4\overline{\mathcal{B}}_{\frac{4}{3}(0,\frac{1}{2})}$ \\
         & $t^{5}v$ & $Q^2u_i$ & -- & $4$ & $4\overline{\mathcal{B}}_{\frac{5}{3}(0,0)}$ \\
         & $t^4v^{-4}$ & $\trace \phi^2$ & -- & $1$ & $\overline{\mathcal{B}}_{\frac{4}{3}(0,0)}$\\
         & $t^5v^{-2}\chi_{\bm{2}}(y)$ & -- & $Q\trace \phi^2$ & $-1$ & $\overline{\mathcal{B}}_{\frac{5}{3}(0,\frac{1}{2})}$ \\
         & $t^6$ & $Q^2\trace \phi^2$ & $1\times$(stress-tensor multiplet) & $0$ & $\overline{\mathcal{B}}_{2(0,0)} + \widehat{\mathcal{C}}_{(0,0)}$ \\
         & $t^6v^{-6}$ & $u_i u_j$, $\trace \phi^3$ & -- & $11$ & $11\overline{\mathcal{B}}_{2(0,0)}$ \\\midrule
        \multirow{3}{*}{$\widehat{I}_{(2,2,2)}$} & $t^4$ & $\trace \mu_i \mu_{j \neq i}$, $Q^2u_i$ & -- & $6$ & $6\overline{\mathcal{B}}_{\frac{4}{3}(0,0)}$ \\
         & $t^5\chi_{\bm{2}}(y)$ & -- & $Qu_i$ & $-3$ & $3\overline{\mathcal{B}}_{\frac{5}{3}(0,\frac{1}{2})}$ \\
         & \multirow{2}{*}{$t^6$} & $u_i$, $\trace \mu_1\mu_2\mu_3$ & \multirow{2}{*}{$2\times$(flavor current)}  & \multirow{2}{*}{$3$} & \multirow{2}{*}{$5\overline{\mathcal{B}}_{2(0,0)} + 2\widehat{\mathcal{C}}_{(0,0)}$} \\
         & & $\trace \mu_1 \mu_3 \mu_2$ & & & \\\midrule
        \multirow{3}{*}{$\widehat{I}_{(2,2,2,2)}$} & $t^{9/2}$ & $u_i$, $Q^2u_i$ & -- & $8$ & $8\overline{\mathcal{B}}_{\frac{3}{2}(0,0)}$ \\
         & $t^{9/2}\chi_{\bm{2}}(y)$ & -- & $Qu_i$  & $-4$ & $4\overline{\mathcal{B}}_{\frac{3}{2}(0,\frac{1}{2})}$ \\
         & $t^6$ & $\trace \mu_i \mu_{j \neq i}$ & $3\times$(flavor current) & $3$ & $6\overline{\mathcal{B}}_{2(0,0)} + 3\widehat{\mathcal{C}}_{(0,0)}$ \\\midrule
        \multirow{4}{*}{$\widehat{I}_{(2,2,2,2,2)}$} & $t^{18/5}$ & $u_i$ & -- & $5$ & $5\overline{\mathcal{B}}_{\frac{6}{5}(0,0)}$\\
         & $t^{21/5}\chi_{\bm{2}}(y)$ & -- & $Qu_i$  & $-5$ & $5\overline{\mathcal{B}}_{\frac{7}{5}(0,\frac{1}{2})}$ \\
         & $t^{24/5}$ & $Q^2u_i$ & -- & $5$ & $5\overline{\mathcal{B}}_{\frac{8}{5}(0,0)}$ \\
         & $t^6$ & -- & $4\times$(flavor current) & $-4$ & $4\widehat{\mathcal{C}}_{(0,0)}$\\\midrule
        \multirow{4}{*}{$\widehat{I}_{(2,2,2,2,2,2)}$} & $t^{3}$ & $u_i$ & -- & $6$ & $6\overline{\mathcal{B}}_{1(0,0)}$ \\
         & $t^{4}\chi_{\bm{2}}(y)$ & -- & $Qu_i$  & $-6$ & $6\overline{\mathcal{B}}_{\frac{4}{3}(0,\frac{1}{2})}$ \\
         & $t^{5}$ & $Q^2u_i$ & -- & $6$ & $6\overline{\mathcal{B}}_{\frac{5}{3}(0,0)}$ \\
         & $t^6$ & $u_i u_j$ & $5\times$(flavor current) & $16$ & $21\overline{\mathcal{B}}_{2(0,0)} + 5\widehat{\mathcal{C}}_{(0,0)}$ \\\midrule
        \multirow{7}{*}{$\widehat{I}_{(2,2),\bm{8}}^{n_a=1}$} & $t^{2.5359}$ & $\trace \phi^2$ & -- & $1$ & $\overline{\mathcal{B}}_{0.8453(0,0)}$ \\
         & $t^{3.80385}$ & $u_i$, $\trace \phi^3$ & -- & $3$ & $3\overline{\mathcal{B}}_{1.26795(0,0)}$ \\
         & $t^{4.26795}\chi_{\bm{2}}(y)$ & -- & $Qu_i$, $\trace \lambda \phi$ & $-3$ & $3\overline{\mathcal{B}}_{1.42265(0,\frac{1}{2})}$ \\
         & $t^{4.73205}$ & $\trace\mu_i\phi$, $Q^2u_i$ & -- & $4$ & $4\overline{\mathcal{B}}_{1.57735(0,0)}$ \\
         & $t^{5.0718}$ & $\left(\trace \phi^2\right)^2$ & -- & $1$ & $\overline{\mathcal{B}}_{1.6906(0,0)}$ \\
         & $t^{5.5359}\chi_{\bm{2}}(y)$ & -- & $\trace \lambda \phi^2$ & $-1$ & $\overline{\mathcal{B}}_{1.8453(0,\frac{1}{2})}$ \\
         & $t^6$ & $\trace\mu_i\phi^2$ & $2\times$(flavor current) & $0$ & $2\overline{\mathcal{B}}_{2(0,0)} + 2\widehat{\mathcal{C}}_{(0,0)}$ \\\bottomrule
    \end{tabular}
    \end{threeparttable}
    \caption{We present relevant and marginal contributions to the indices associated to theories achieved via various ways of gauging copies of $\mathcal{D}_2(SU(3))$. For an explanation of the notation for the operators, see the main text. The positive/negative columns summarize the operators that contribute either positively or negatively to the index, and we sum those contributions in the final column, which is the the coefficient of the term in the index. The ``flavor current'' at order $t^6$ refers to the leading order contribution from the supermultiplet containing the flavor current; this contribution comes from a fermionic component.}
    \label{tbl:2s}
\end{table}

\begin{table}[H]
    \centering
    \footnotesize
    \renewcommand{\arraystretch}{1.2}
    \begin{threeparttable}
    \begin{tabular}{cccccc}
    \toprule
        \multirow{2}{*}{Index} & \multirow{2}{*}{Term} & \multirow{2}{*}{Positive} & \multirow{2}{*}{Negative} & \multirow{2}{*}{Total} & $\mathcal{N}=1$ \\
         & & & & & Multiplets \\\midrule
        \multirow{7}{*}{$\widehat{I}_{\widehat{E}_6(SU(2))}$} & $t^{8/3}v^{-8/3}$ & $u_i$ & -- & $3$ &$3\overline{\mathcal{B}}_{\frac{8}{9}(0,0)}$ \\
         & $t^{11/3}v^{-2/3}\chi_{\bm{2}}(y)$ & -- & $Qu_i$ & $-3$ & $3\overline{\mathcal{B}}_{\frac{11}{9}(0,\frac{1}{2})}$ \\
         & $t^{14/3}v^{4/3}$ & $Q^2u_i$ & -- & $3$ & $3\overline{\mathcal{B}}_{\frac{14}{9}(0,0)}$ \\
         & $t^{16/3}v^{-16/3}$ & $u_i u_j$ & -- & $6$ & $6\overline{\mathcal{B}}_{\frac{8}{3}(0,0)}$ \\
         & $t^4v^{-4}$ & $\trace \phi^2$ & -- & $1$ & $\overline{\mathcal{B}}_{\frac{4}{3}(0,0)}$ \\
         & $t^5v^{-2}\chi_{\bm{2}}(y)$ & -- & $Q\trace \phi^2$ & $-1$ & $\overline{\mathcal{B}}_{\frac{5}{3}(0,\frac{1}{2})}$ \\
         & $t^6$ & $Q^2\trace \phi^2$ & $1\times$(stress-tensor multiplet) & $0$ & $\overline{\mathcal{B}}_{2(0,0)} + \widehat{\mathcal{C}}_{(0,0)}$\\\midrule
        \multirow{4}{*}{$\widehat{I}_{(3,3)}$} & $t^3$ & $Q^2u_i$, $\trace \mu_1 \mu_2$ & -- & $3$ & $3\overline{\mathcal{B}}_{1(0,0)}$ \\
         & $t^{9/2}\chi_{\bm{2}}(y)$ & -- & $Qu_i$ & $-2$ & $2\overline{\mathcal{B}}_{\frac{3}{2}(0,\frac{1}{2})}$ \\
         & \multirow{2}{*}{$t^6$} & $u_i$, $(\trace \mu_1 \mu_2)^2$ & \multirow{2}{*}{$1\times$(flavor current)}  & \multirow{2}{*}{$3$} & \multirow{2}{*}{$4\overline{\mathcal{B}}_{2(0,0)} + \widehat{\mathcal{C}}_{(0,0)}$} \\
         & & $(Q^2u_{1})(Q^2u_{2})$ & & \\\midrule
        \multirow{3}{*}{$\widehat{I}_{(3,3,3)}$} & $t^4$ & $u_i$, $Q^2u_i$ & -- & $6$ & $6\overline{\mathcal{B}}_{\frac{4}{3}(0,0)}$ \\
         & $t^4\chi_{\bm{2}}(y)$ & -- & $Qu_i$ & $-3$ & $3\overline{\mathcal{B}}_{\frac{4}{3}(0,\frac{1}{2})}$ \\
         & $t^6$ & $\trace \mu_i \mu_{j\neq i}$ & $2\times$(flavor current)  & $1$ & $3\overline{\mathcal{B}}_{2(0,0)} + 2\widehat{\mathcal{C}}_{(0,0)}$ \\\midrule
        \multirow{4}{*}{$\widehat{I}_{(3,3,3,3)}$} & $t^3$ & $u_i$ & -- & $4$ & $4\overline{\mathcal{B}}_{1(0,0)}$ \\
         & $t^{15/4}\chi_{\bm{2}}(y)$ & -- & $Qu_i$ & $-4$ & $4\overline{\mathcal{B}}_{\frac{5}{4}(0,\frac{1}{2})}$ \\
         & $t^{9/2}$ & $Q^2u_i$ & -- & $4$ & $4\overline{\mathcal{B}}_{\frac{3}{2}(0,0)}$ \\
         & $t^6$ & $u_i u_j$ & $3\times$(flavor current) & $7$ & $10\overline{\mathcal{B}}_{2(0,0)} + 3\widehat{\mathcal{C}}_{(0,0)}$ \\\midrule
        \multirow{8}{*}{$\widehat{I}_{(3),\bm{3},\bm{3}}^{n_a=2}$} & $t^{2.74273}$ & $u$ & -- & $1$ & $\overline{\mathcal{B}}_{0.91424(0,0)}$ \\
         & $t^{3.68568}\chi_{\bm{2}}(y)$ & -- & $Qu$ & $-1$ & $\overline{\mathcal{B}}_{1.22856(0,\frac{1}{2})}$ \\
         & $t^{4.62864}$ & $Q^2u$ & -- & $1$ & $3\overline{\mathcal{B}}_{1.54288(0,0)}$ \\
         & $t^{3.68568}$ & $\trace \phi_i \phi_{j}$ & -- & $3$ & $\overline{\mathcal{B}}_{1.22856(0,0)}$\\
         & $t^{4.84284}\chi_{\bm{2}}(y)$ & -- & $\trace\lambda\phi_i$ & $-2$ & $2\overline{\mathcal{B}}_{1.61428(0,\frac{1}{2})}$ \\
         & $t^{5.48545}$ & $u^2$ & -- & $1$ & $\overline{\mathcal{B}}_{1.82848(0,0)}$ \\
         & $t^{5.7858}$ & $\trace \mu \phi_i$ & -- & $2$ & $2\overline{\mathcal{B}}_{1.9286(0,0)}$\\
         & $t^6$ & -- & $4\times$(flavor current) & $-4$ & $4\widehat{\mathcal{C}}_{(0,0)}$ \\\bottomrule
    \end{tabular}
    \end{threeparttable}
    \caption{We write the relevant and marginal contributions to the indices associated to the theories obtained via various ways of gauging copies of $\mathcal{D}_3(SU(2))$. The notation is as described in Section \ref{sec:summary}. Again, the positive/negative columns summarize the operators that contribute positively/negatively to the index. The total column is the sum of the positive and negative contributions and provides the coefficient of the associated term in the index. The ``flavor current'' at order $t^6$ refers to the leading order contribution from the supermultiplet containing the flavor current; this contribution comes from a fermionic component.}
    \label{tbl:3s}
\end{table}

\begin{table}[H]
    \centering
    \footnotesize
    \renewcommand{\arraystretch}{1.2}
    \begin{threeparttable}
    \begin{tabular}{cccccc}
    \toprule
        \multirow{2}{*}{Index} & \multirow{2}{*}{Term} & \multirow{2}{*}{Positive} & \multirow{2}{*}{Negative} & \multirow{2}{*}{Total} & $\mathcal{N}=1$ \\
         & & & & & Multiplets \\\midrule
        \multirow{9}{*}{$\widehat{I}_{(3,3,5)}$} 
         & $t^{3.05814}$ & $u_3$ & -- & $1$ & $\overline{\mathcal{B}}_{1.01938(0,0)}$ \\
         & $t^{3.50969}\chi_{\bm{2}}(y)$ & -- & $Qu_3$ & $-1$ & $\overline{\mathcal{B}}_{3.50970(0,\frac{1}{2})}$ \\
         & $t^{3.96124}$ & $u_1$, $u_2$, $Q^2u_3$ & -- & $3$ & $3\overline{\mathcal{B}}_{1.3204(0,0)}$ \\
         & $t^{3.99031}\chi_{\bm{2}}(y)$ & -- & $Qu_1$, $Qu_2$ & $-2$ & $2\overline{\mathcal{B}}_{1.3301(0,\frac{1}{2})}$ \\
         & $t^{4.01938}$ & $Q^2u_1$, $Q^2u_2$ & -- & $2$ & $2\overline{\mathcal{B}}_{1.33979(0,0)}$ \\
         & $t^{4.07752}$ & $v_3$ & -- & $1$ & $\overline{\mathcal{B}}_{1.35917(0,0)}$ \\
         & $t^{4.52907}\chi_{\bm{2}}(y)$ & -- & $Qv_3$ & $-1$ & $\overline{\mathcal{B}}_{1.50969(0,\frac{1}{2})}$ \\
         & $t^{4.98062}$ & $Q^2v_3$ & -- & $1$ & $\overline{\mathcal{B}}_{1.6602(0,0)}$ \\
         & $t^6$ & -- & $2\times$(flavor current) & $-2$ & $2\widehat{\mathcal{C}}_{(0,0)}$ \\\midrule
        \multirow{5}{*}{$\widehat{I}_{(5,5)}$} 
         & $t^{3}$ & $Q^2u_i$ & -- & $2$ & $2\overline{\mathcal{B}}_{1(0,0)}$ \\
         & $t^{15/4}\chi_{\bm{2}}(y)$ & -- & $Qu_i$ & $-2$ & $2\overline{\mathcal{B}}_{\frac{5}{4}(0,\frac{1}{2})}$ \\
         & $t^{9/2}$ & $u_i$, $Q^2v_i$, $\trace \mu_1 \mu_2$ & -- & $5$ & $5\overline{\mathcal{B}}_{\frac{3}{2}(0,0)}$ \\
         & $t^{21/4}$ & -- & $Q^2v_i$ & $-2$ & $2\overline{\mathcal{B}}_{\frac{7}{4}(0,\frac{1}{2})}$ \\
         & $t^6$ & $v_i$, $Q^2u_1 Q^2u_2$ & $1\times$(flavor current) & $2$ & $3\overline{\mathcal{B}}_{2(0,0)} + \widehat{\mathcal{C}}_{(0,0)}$
         \\\midrule
        \multirow{13}{*}{$\widehat{I}_{(5),\bm{3},\bm{3}}^{n_a=2}$} & $t^{2.42423}$ & $u$ & -- & $1$ & $\overline{\mathcal{B}}_{0.80808(0,0)}$ \\
        & $t^{3.23231}$ & $v$ & -- & $1$ & $\overline{\mathcal{B}}_{1.0774(0,0)}$ \\
        & $t^{3.40404}\chi_{\bm{2}}(y)$ & -- & $Qu$ & $-1$ & $\overline{\mathcal{B}}_{1.13468(0,\frac{1}{2})}$ \\
        & $t^{3.80808}$ & $\trace \phi_1^2$, $\trace \phi_1\phi_2$, $\trace\phi_2^2$ & -- & $3$ & $3\overline{\mathcal{B}}_{1.26936(0,0)}$ \\
        & $t^{4.21212}\chi_{\bm{2}}$ & -- & $Qv$ & $-1$ & $\overline{\mathcal{B}}_{1.40404(0,\frac{1}{2})}$ \\
        & $t^{4.38384}$ & $Q^2u$ & -- & $1$ & $\overline{\mathcal{B}}_{1.46128(0,0)}$ \\
        & $t^{4.84847}$ & $u^2$ & -- & $1$ & $\overline{\mathcal{B}}_{1.61616(0,0)}$ \\
        & $t^{4.90404}\chi_{\bm{2}}(y)$ & -- & $\trace \lambda\phi_1$, $\trace \lambda\phi_2$ & $-2$ & $2\overline{\mathcal{B}}_{1.63468(0,\frac{1}{2})}$ \\
        & $t^{5.19192}$ & $Q^2v$ & -- & $1$ & $\overline{\mathcal{B}}_{1.73064 (0,0)}$ \\
        & $t^{5.65654}$ & $uv$ & -- & $1$ & $\overline{\mathcal{B}}_{1.8855(0,0)}$ \\
        & $t^{5.82827}\chi_{\bm{2}}(y)$ & -- & $uQu$ & $-1$ & $\overline{\mathcal{B}}_{1.9428(0,\frac{1}{2})}$ \\
        & $t^{5.88384}$ & $\trace\mu\phi_1$, $\trace\mu\phi_2$ & -- & $2$ & $2\overline{\mathcal{B}}_{1.96128 (0,0)}$ \\
        & $t^6$ & -- & $4\times$(flavor current) & $-4$ & $4\widehat{\mathcal{C}}_{(0,0)}$
        \\ \bottomrule
    \end{tabular}
    \end{threeparttable}
    \caption{In this table, we summarize the superconformal indices, and the associated operator spectroscopy, for the gaugings discussed in Sections \ref{sec:indexirrational}, \ref{sec:1D5}, and \ref{sec:noDp}. Again, the positive/negative columns summarize the operators that contribute positively/negatively to the index. The total column is the sum of the positive and negative contributions and provides the coefficient of the associated term in the index. The ``flavor current'' at order $t^6$ refers to the leading order contribution from the supermultiplet containing the flavor current; this contribution comes from a fermionic component.}
    \label{tbl:5s}
\end{table}

\section*{Acknowledgements}

We thank Richard Derryberry for a helpful discussion. M.J.K.~and C.L.~thank the KAIX program of KAIST for support during the final stage of this work. M.J.K.~is supported by Sherman Fairchild Postdoctoral Fellowship and the U.S.~Department of Energy, Office of Science, Office of High Energy Physics, under Award Number DE-SC0011632. C.L.~acknowledges support from DESY (Hamburg, Germany), a member of the Helmholtz Association HGF. K.H.L.~and J.S.~are partly supported by the NRF grant NRF-2020R1C1C1007591. The work of J.S.~is also supported by POSCO Science Fellowship of POSCO TJ Park Foundation and a Start-up Research Grant for new faculty provided by KAIST. 

\appendix

\section{Superconformal indices with flavor fugacities}\label{app:indicesplusplus}

In this appendix, we list the superconformal indices that were worked out throughout this paper, in a refined way where all of the fugacities for the flavor symmetries are turned on. The expressions tend to be rather cumbersome, as the flavor symmetry is generically just $U(1)^N$, for some $N$, however we write them here for both completeness and future reference. We emphasize that throughout this appendix, we are considering the flavor-fugacity-refined version of the reduced superconformal index, defined as in equation \eqref{eqn:reducedindex}. 

The refined and reduced index of three $\mathcal{D}_2(SU(3))$ gauged together by an $\mathcal{N}=1$ $SU(3)$ vector multiplet is
\begin{align}
    \begin{split}
        \widehat{I}_{(2,2,2)}& =t^4\left(v_1^{2}+v_1^{-1}+v_1^{-2}v_2^{2}+v_2+v_1v_2^{-1}+v_2^{-2}\right)-t^5\chi_{\bm{2}}(y)\left(v_1+v_1^{-1}v_2+v_2^{-1}\right)\\
        &\quad +t^6\left(-2+v_1^{3}+v_1^{-3}v_2^{3}+v_2^{-3}\right)-t^7\chi_{\bf{2}}(y)(v_1^{-2}+v_1^2v_2^{-2}+v_2^2)\\
        &+t^8(v_1^{-4}+2v_1^{-1}+v_1^2+v_1^4v_2^{-4}+v_2^{-2}+2v_1v_2^{-1}+2v_2+v_1^{-2}v_2^2+v_2^4)+O\left(t^9\right) \,.
    \end{split}
    \label{eqn:refinedI222}
\end{align}
In comparison with the unrefined index in equation \eqref{eqn:ind222}, we have here two fugacities $v_1$ and $v_2$ associated to the two $U(1)$ flavor symmetries generated by $\mathcal{F}_2-\mathcal{F}_1$ and $\mathcal{F}_3-\mathcal{F}_2$. We can see that equation \eqref{eqn:ind222} is recovered when we take $v_1 = v_2 = 1$, as required. 
Next, we write down the reduced index of four $\mathcal{D}_2(SU(3))$ theories glued by $\mathcal{N}=1$ gauging, with the three fugacities $v_i$ for the three $U(1)$ flavor symmetries $\mathcal{F}_{i+1}-\mathcal{F}_{i}$. The refined index is 
\begin{align}
    \begin{split}
        \widehat{I}_{(2,2,2,2)}&=t^{\frac{9}{2}}\left(\left(v_1^{-1}+v_1^3+v_1v_2^{-1}+v_1^{-3}v_2^3+v_3^{-3}+v_2v_3^{-1}+v_3+v_2^{-3}v_3^3\right)\right. \\
        &\qquad\qquad \left. -\chi_{\bm{2}}(y)\left(v_1+v_1^{-1}v_2+v_2^{-1}v_3+v_3^{-1}\right)\right)\\
        &\quad +t^6\left(-3+v_2^{-2}+v_2^2+v_1^2v_3^{-2}+v_1^{-2}v_2^2v_3^{-2}+v_1^{-2}v_3^2+v_1^2v_2^{-2}v_3^2\right)+O\left(t^9\right) \,.
    \end{split}
\end{align}
Similarly, the refined index for five $\mathcal{D}_2(SU(3))$ glued together involves four fugacities $v_{i}$ associated to the four non-anomalous $U(1)$ flavor symmetries, $\mathcal{F}_{i+1}-\mathcal{F}_{i}$; we find
\begin{align}
\begin{split}
    \widehat{I}_{(2,2,2,2,2)}&=t^{\frac{18}{5}}\left(v_1^3+v_1^{-3}v_2^3+v_2^{-3}v_3^3+v_3^{-3}v_4^3+v_4^{-3}\right)\\
&\quad-t^{\frac{21}{5}}\chi_{\bm{2}}(y)\left(v_1+v_1^{-1}v_2+v_2^{-1}v_3+v_3^{-1}v_4+v_4^{-1}\right)\\
    &\quad +t^{\frac{24}{5}}\left(v_1^{-1}+v_1v_2^{-1}+v_2v_3^{-1}+v_3v_4^{-1}+v_4\right)-4t^6+O\left(t^{\frac{36}{5}}\right) \,.
\end{split}
\end{align}
Finally, the refined index of the conformal $\mathcal{N}=1$ gauging of six copies of $\mathcal{D}_2(SU(3))$ is
\begin{align}
\begin{split}
\widehat{I}_{(2,2,2,2,2,2)}&=t^3\left(v_1^3+v_1^{-3}v_2^3+v_2^{-3}v_3^3+v_3^{-3}v_4^3+v_4^{-3}v_5^3+v_5^{-3}\right)\\
&\quad -t^4\chi_{\bm{2}}(y)\left(v_1+v_1^{-1}v_2+v_2^{-1}v_3+v_3^{-1}v_4+v_4^{-1}v_5+v_5^{-1}\right)\\
&\quad +t^5\left(v_1^{-1}+v_1v_2^{-1}+v_2v_3^{-1}+v_3v_4^{-1}+v_4v_5^{-1}+v_5\right)\\
&\quad +t^6(-5+v_1^6+v_2^{3}+v_1^{-6}v_2^6+v_1^{-3}v_3^3+v_1^3v_2^{-3}v_3^3+v_2^{-6}v_3^6+v_4^{-3}\\
&\qquad+v_2^{-3}v_4^3+v_1^3v_3^{-3}v_4^3+v_1^{-3}v_2^3v_3^{-3}v_4^3+v_3^{-6}v_4^6+v_5^{-6}+v_1^3v_5^{-3}\\
&\qquad+v_1^{-3}v_2^3v_5^{-3}+v_2^{-3}v_3^3v_5^{-3}+v_3^{-3}v_4^3v_5^{-3}+v_3^{-3}v_5^3+v_1^3v_4^{-3}v_5^3\\
&\qquad+v_1^{-3}v_2^3v_4^{-3}v_5^3+v_2^{-3}v_3^3v_4^{-3}v_5^3+v_4^{-6}v_5^6)+O\left(t^7\right) \,.
\end{split}
\end{align}
The fugacities $v_{i}$ are again associated to the five $U(1)$ flavor symmetries, which are generated by $\mathcal{F}_{i+1}-\mathcal{F}_{i}$.

Next, we turn our attention to the refined indices involving the gauging of $\mathcal{D}_3(SU(2))$ theories.
The refined index of two $\mathcal{D}_3(SU(2))$ gauged together, with a fugacity $v$ for the flavor $U(1)$ generated by $\mathcal{F}_{2}-\mathcal{F}_{1}$, is
\begin{align}
\label{eq:I33ref}
\begin{split}
\widehat{I}_{(3,3)}&=t^3\left(1+v^{\frac{4}{3}}+v^{-\frac{4}{3}}\right)-t^{\frac{9}{2}}\chi_{\bm{2}}(y)\left(v^{\frac{2}{3}}+v^{-\frac{2}{3}}\right)+t^6\left(1+v^{\frac{8}{3}}+v^{-\frac{8}{3}}\right)\\
&-t^{\frac{15}{2}}\chi_{\bm{2}}(y)\left(v^2+v^{-2}\right)+t^9\left(1+v^4+v^{-4}+\chi_{\bm{2}}(y)\left(v^{\frac{4}{3}}+v^{-\frac{4}{3}}\right)\right)+O\left(t^\frac{21}{2}\right) \,.
\end{split}
\end{align}
Similarly, the refined index of three $\mathcal{D}_3(SU(2))$ gauged together via the diagonal of the $SU(2)$ flavor symmetries is
\begin{align}
\begin{split}
\widehat{I}_{(3,3,3)}&=t^4\left(v_1^{-\frac{4}{3}}+v_1^{\frac{8}{3}}+v_2^{-\frac{8}{3}}+v_1^{\frac{4}{3}}v_2^{-\frac{4}{3}}+v_2^{\frac{4}{3}}+v_1^{-\frac{8}{3}}v_2^{\frac{8}{3}}-\chi_{\bm{2}}(y)\left(v_1^{\frac{2}{3}}+v_2^{-\frac{2}{3}}+v_1^{-\frac{2}{3}}v_2^{\frac{2}{3}}\right)\right)\\
&\quad +t^6\left(v_1^2+v_2^{-2}+v_1^{-2}v_2^2-2\right)+O\left(t^8\right) \,,
\end{split}
\end{align}
where, again, we have introduced two fugacities $v_{i}$ for the two non-anomalous $U(1)$ flavor symmetries generated by $\mathcal{F}_{i+1}-\mathcal{F}_{i}$.
When four copies of the $\mathcal{D}_3(SU(2))$ SCFT are $(\mathcal{N}=1)$-gauged together via their common flavor symmetry, we find that the refined index, with three fugacities $v_{i}$ standing for the three $U(1)$ flavor symmetries associated to $\mathcal{F}_{i+1}-\mathcal{F}_{i}$, is
\begin{align}
\begin{split}
\widehat{I}_{(3,3,3,3)}&=t^3\left(v_1^{\frac{8}{3}}+v_1^{-\frac{8}{3}}v_2^{\frac{8}{3}}+v_2^{-\frac{8}{3}}v_3^{\frac{8}{3}}+v_3^{-\frac{8}{3}}\right)-t^{\frac{15}{4}}\chi_{\bm{2}}(y)\left(v_1^{\frac{2}{3}}+v_1^{-\frac{2}{3}}v_2^{\frac{2}{3}}+v_2^{-\frac{2}{3}}v_3^{\frac{2}{3}}+v_3^{-\frac{2}{3}}\right)\\
&\quad + t^{\frac{9}{2}}\left(v_1^{-\frac{4}{3}}+v_1^{\frac{4}{3}}v_2^{-\frac{4}{3}}+v_2^{\frac{4}{3}}v_3^{-\frac{4}{3}}+v_3^{\frac{4}{3}}\right)+t^6\Big(-3+v_1^{\frac{16}{3}}+v_2^{-\frac{8}{3}}+v_2^{\frac{8}{3}}+v_1^{-\frac{16}{3}}v_2^{\frac{16}{3}}\\
&\qquad+v_3^{-\frac{16}{3}}+v_1^{\frac{8}{3}}v_3^{-\frac{8}{3}}+v_1^{-\frac{8}{3}}v_2^{\frac{8}{3}}v_3^{-\frac{8}{3}}+v_1^{-\frac{8}{3}}v_3^{\frac{8}{3}}+v_1^{\frac{8}{3}}v_2^{-\frac{8}{3}}v_3^{\frac{8}{3}}+v_2^{-\frac{16}{3}}v_3^{\frac{16}{3}}\Big)\\
&\quad+O\left(t^{\frac{27}{4}}\right) \,.
\end{split}
\end{align}

We have now written the flavor-fugacity-refined reduced superconformal indices for all the 4d $\mathcal{N}=1$ SCFTs arising from either asymptotically-free or conformal gaugings of the diagonal of the flavor symmetry of a collection of either $\mathcal{D}_2(SU(3))$ or $\mathcal{D}_3(SU(2))$ theories. We now include the refined indices for some more sporadic examples of the $\mathcal{N}=1$ SCFTs with $a=c$ that were determined in \cite{Kang:2021ccs}.
The refined index for the infrared SCFT arising from the $\mathcal{N}=1$ gauging of the $SU(2)$ flavor symmetries of two copies of $\mathcal{D}_3(SU(2))$ and one copy of $\mathcal{D}_5(SU(2))$ is
\begin{align}
\begin{split}
\widehat{I}_{(3,3,5)}&=t^{3.05814}v_2^{12}-t^{3.50969}\chi_{\bm{2}}(y)v_2^2+t^{3.96124}\left(v_1^{\frac{8}{3}}+v_1^{-\frac{8}{3}}+1\right)v_2^{-8}\\
&\quad-t^{3.99031}\chi_{\bm{2}}(y)\left(v_1^{\frac{2}{3}}+v_1^{-\frac{2}{3}}\right)v_2^{-2} +t^{4.01938}\left(v_1^{\frac{4}{3}}+v_1^{-\frac{4}{3}}\right)v_2^4+t^{4.07752}v_2^{16}\\
&\quad-t^{4.52907}\chi_{\bm{2}}(y)v_2^6+t^{4.98062}v_2^{-4}-2t^6+O\left(t^{6.05814}\right) \,.
\end{split}
\end{align}
There are two $U(1)$ flavor symmetries which do not have an ABJ anomaly, and to which we associated the fugacities $v_1$ and $v_2$. These correspond to the $U(1)$s generated by, respectively, $-\mathcal{F}_1+\mathcal{F}_2$ and $3\mathcal{F}_1+3\mathcal{F}_2-5\mathcal{F}_3$; here $\mathcal{F}_{1,2}$ are the flavor $U(1)$s arising from the two $D_3(SU(2))$s and $\mathcal{F}_3$ is the flavor $U(1)$ coming from the $\mathcal{D}_5(SU(2))$.
We can also consider the refined index of the theory arising from two $\mathcal{D}_5(SU(2))$ gauged together, where there is one flavor fugacity $v$ standing for the anomaly-free $U(1)$ symmetry generated by $-\mathcal{F}_{1}+\mathcal{F}_2$. This index is
\begin{align}
\begin{split}
\widehat{I}_{(5,5)}&=t^3\left(v^{\frac{8}{5}}+v^{-\frac{8}{5}}\right)-t^{\frac{15}{4}}\chi_{\bm{2}}(y)\left(v^{\frac{2}{5}}+v^{-\frac{2}{5}}\right)+t^{\frac{9}{2}}\left(1+v^{\frac{12}{5}}+v^{\frac{4}{5}}+v^{-\frac{4}{5}}+v^{-\frac{12}{5}}\right)\\
&\quad -t^{\frac{21}{4}}\chi_{\bm{2}}(y)\left(v^{\frac{6}{5}}+v^{-\frac{6}{5}}\right)+t^6\left(v^{\frac{16}{5}}+v^{-\frac{16}{5}}\right)+O\left(t^{\frac{27}{4}}\right) \,.
\end{split}
\end{align}
Next, we include an example where we are not only gauging a collection of Argyres--Douglas theories, by also where we include additional chiral matter multiplets charged under the introduced gauge node. In particular, the refined reduced index for the theory obained via gauging two $\mathcal{D}_2(SU(3))$ theories together by $\mathcal{N}=1$ gauging with one additional adjoint chiral multiplet $\phi$ is
\begin{align}
\begin{split}
\widehat{I}^{n_a=1}_{(2,2),\bm{8}}&=t^{2.5359}v_2^{-2}+t^{3.80385}\left(v_1^3+1+v_1^{-3}\right)v_2^{-3}-t^{4.26795}\chi_{\bm{2}}(y)\left(1+v_1+v_1^{-1}\right)v_2^{-1}\\
&\quad +t^{4.73205}\left(v_1^2+v_1+v_1^{-1}+v_1^{-2}\right)v_2+t^{5.0718}v_2^{-4}-t^{5.5359}\chi_{\bm{2}}(y)v_2^{-2}\\
&\quad +t^6\left(v_1^2+v_1^{-2}-2\right)+O\left(t^{6.33975}\right) \,.
\end{split}
\end{align}
The two fugacities $v_{1}$ and $v_2$ are for the two flavor $U(1)$ generated by $-\mathcal{F}_1+\mathcal{F}_2$ and $\mathcal{F}_1+\mathcal{F}_2-T$, where $T$ is the classical $U(1)$ symmetry that rotates the phase of $\phi$.
As another example involving adjoint-valued chiral multiplets, we consider the refined index for one $\mathcal{D}_3(SU(2))$ where the $SU(2)$ flavor symmetry is gauged together with two adjoint chirals $\phi_{1}$ and $\phi_2$. There is an  $SU(2)\times U(1)$ flavor symmetry; the $SU(2)$ rotates the two chirals and the remaining non-anomalous $U(1)$ is $T-3\mathcal{F}$, where the $T$ is the generator of the $U(1)$ factor in the $U(2)$ that classically rotates the two adjoint chiral multiplets. The fugacity for the Cartan of $SU(2)$ is $v_1$ and that for the $U(1)$ is $v_2$. Altogether, the refined reduced index is
\begin{align}
\begin{split}
\widehat{I}^{n_a=2}_{(3),\bm{3},\bm{3}}&=t^{2.74273}v_2^8+t^{3.68568}\left(\chi_{\mathfrak{su}_2,\bm{3}}(v_1)v_2^2-\chi_{\bm{2}}(y)v_2^2\right)+t^{4.62864}v_2^{-4}\\
&\quad-t^{4.84284}\chi_{\bm{2}}(y)\chi_{\mathfrak{su}_2,\bm{2}}(v_1)v_2 +t^{5.48545}v_2^{16}+t^{5.78580}\chi_{\mathfrak{su}_2,\bm{2}}(v_1)v_2^{-5}\\
&\quad-t^6\left(\chi_{\mathfrak{su}_2,\bm{3}}(v_1)+1\right)+O\left(t^{6.42841}\right) \,.
\end{split}
\end{align}
From all of these refined indices we find that the charges of the operators under the $U(1)$ flavor symmetries, as read off from the fugacities, matches the identification of operators using operator spectroscopy.
Finally, we consider the refined index for one $\mathcal{D}_5(SU(2))$ whose flavor $SU(2)$ is gauged and coupled to two adjoint chirals $\phi_1$ and $\phi_2$. Similar to the previous example of $\mathcal{D}_3(SU(2))$ gauged with two adjoint chirals, this theory also has $SU(2)\times U(1)$; two adjoint chirals are rotated into each other by an $SU(2)$ flavor symmetry, and there is also a non-anomalous $U(1)$ that is $5T-4\mathcal{F}$ where $T$ rotates the phases of adjoint chirals. We turn on fugacities $v_1$ and $v_2$, which are associated to the $SU(2)$ and $U(1)$ flavor symmetries, respectively. The refined index of this theory is given by
\begin{align}
\begin{split}
    \widehat{I}^{n_a=2}_{(5),\bm{3},\bm{3}}=\ &
    t^{2.42423}v_2^{\frac{48}{5}}+t^{3.23231}v_2^{\frac{64}{5}}-t^{3.40404}\chi_{\bm{2}}(y)v_2^{\frac{8}{5}}+t^{3.80808}\chi_{\mathfrak{su}_2,\bm{3}}(v_1)v_2^{10}\\
    &-t^{4.21212}\chi_{\bm{2}}(y)v_2^{\frac{24}{5}}+t^{4.38384}v_2^{-\frac{32}{5}}+t^{4.84847}v_2^{\frac{96}{5}}-t^{4.90404}\chi_{\bm{2}}(y)\chi_{\mathfrak{su}_2,\bm{2}}(v_1)v_2^5\\
    &+t^{5.19192}v_2^{-\frac{16}{5}}-t^{5.65654}v_2^{\frac{112}{5}}-t^{5.82827}\chi_{\bm{2}}(y)v_2^{\frac{56}{5}}+t^{5.88384}\chi_{\mathfrak{su}_2,\bm{2}}(v_1)v_2^{-3}\\
    &-t^6\left(\chi_{\mathfrak{su}_2,\bm{2}}(v_1)+1\right)+O\left(t^{6.23231}\right)\,.
\end{split}
\end{align}

\bibliography{references}{}
\bibliographystyle{sortedbutpretty}

\end{document}